\documentclass[12pt]{article}
\usepackage{makeidx}
\usepackage{epsfig}
\usepackage{arydshln}
\usepackage{float}
\usepackage{amscd}
\usepackage{amsmath, amsthm}
\usepackage{amssymb}
\usepackage{enumitem}
\usepackage{color}
\usepackage{multirow}
\usepackage{fancyheadings}
\usepackage{fancyvrb}
\usepackage{xcolor}
\usepackage{url}
\usepackage{subcaption}

\usepackage{tikz}
\usepackage{overpic}

\newcommand{\newgamma}{\mu}
\newcommand{\coloneqq}{:=}
\newcommand{\eqqcolon}{=:}

\newcommand{\bchi}{{\boldsymbol \chi}}
\newcommand{\bog}{\Gamma}

\newtheorem{lemma}{Lemma}
\newtheorem{theorem}{Theorem}
\newtheorem{remark}{Remark}

\newcommand{\tbtau}{{\boldsymbol \tau}_{\kern-.2em\pi}}
\newcommand{\omitit}[1]{}
\newcommand{\Reyuls}{\mathbb{R}}

\newcommand{\p}{{{\rm I}{\relax\ifmmode\mskip-\thinmuskip\relax\else\kern-.22em\fi}{\bf P}}}
\newcommand{\perpzh}{\perp\kern-.20em^a}

\newcommand{\Plang}{{\p}{\kern-.03em{\rm fortran}}}
\newcommand{\Pfortran}{{\p}{\kern-.03em{\rm fortran}}}

\newcommand{\half}{{\textstyle{1\over 2}}}

\newcommand{\set}[2]{\left\lbrace #1 \; : \; #2 \right\rbrace}

\newcommand{\bo}{{\partial\Omega}}

\newcommand{\du}{{\mathcal{D}}(\uu)}

\newcommand{\dv}{{\mathcal{D}}(\vv)}

\newfloat{code}{ht}{aux}[section]
\floatstyle{boxed}
\floatname{code}{Program}
\restylefloat{code}

\newcommand{\norm}[1]{\Vert{#1}\Vert}

\newcommand{\tbnorm}[1]{\vert\kern-.1em\vert\kern-.1em\vert{\, #1 \,}\vert\kern-.1em\vert\kern-.1em\vert}

\newcommand{\oput}[1]{\rlap{${}{\lower 1.3ex\hbox{${\sim}$}}$}{#1}}

\newcommand{\sdiv}{{{\nabla\cdot} \,}}
\newcommand{\curl}{{{\rm curl} \,}}

\newcommand{\bfz}{{\mathbf 0}}

\newcommand{\uu}{{\mathbf u}}

\newcommand{\vv}{{\mathbf v}}
\newcommand{\ww}{{\mathbf w}}

\newcommand{\gbc}{{\mathbf g}}

\newcommand{\btek}{{\mathbf t}_{e,T}^{\kern.1em\prime}}
\newcommand{\bteke}{{\mathbf t}_{e,T_k}^{\kern.1em\prime}}
\newcommand{\btep}{{\mathbf t}_{e,k}^{\kern.1em\prime}}
\newcommand{\btpj}{{\mathbf t}_j^{\kern.1em\prime}}
\newcommand{\nn}{{\mathbf n}}

\newcommand{\ff}{{\mathbf f}}

\newcommand{\xx}{{\mathbf x}}

\newcommand{\intox}[1]{\int_{\Omega} #1 \, d\xx}

\newcommand{\datekern}{{\relax\kern+.1em}--{\relax\kern-.01em}}

\newcommand{\antiparallel}{\not\kern-0.35em\Vert}

\newcommand{\derdir}[2]{{\frac{\partial{#1}}{\partial #2}}}

\newtheorem{counterex}{Example}

\makeindex
\usepackage{makeidx}

\begin{document}
\title{Verification and Validation of Cylinder Drag: Pressure and Stress Approximations on Curved Boundaries}
\author{Ingeborg G. Gjerde, Simula \\
L. Ridgway Scott, University of Chicago}
\def\thepage{}\maketitle\pagenumbering{arabic}

\begin{abstract}
We study a technique for verification of stress and pressure computations on boundaries in flow simulations.
We utilize existing experiments to provide validation of the simulations. We show that this approach can reveal critical flaws in simulation algorithms. Using the successful computational algorithms, we examine Lamb's model  for cylinder drag at low Reynolds numbers. We comment on a discrepancy observed in an experimental paper, suggesting
that the domain size may be a contributing factor.
Our simulations on suitably large domains confirm Lamb's model.
We highlight a paradox related to imposing Dirichlet (Stokes) boundary
conditions on polygonal approximations of the curved surface using
finite-element methods that are exactly divergence free.
The finite-element simulations provide very poor representations of drag
when the boundary conditions are imposed strongly.
We demonstrate that relaxing the boundary conditions using Nitsche's method
restores high-order approximation.
\end{abstract}

Verification and validation
\cite{adrion1982validation,szabo2021finite,wallace1989software}
are essential steps for assessing software reliability
and its credibility in simulations. In the context of numerical methods for
flow simulation, verification typically concerns the accuracy of the numerical method
and its implementation (are you obtaining the right solution?), while validation
concerns the validity of the model (are you solving the right system?).
In this paper we describe two techniques for verification and validation of drag
in general domains, and show how they uncovered a subtle but serious numerical
issue for exactly divergence-free discretizations.

The verification method we consider uses two ways to compute drag \cite{john1997parallele,ref:refvalcyliftdragVolkerJohn}, involving the volume integral rather than the surface integral, the usual definition. We apply this method to four well-established
finite-element methods for fluid flow, the P2-P0 method \cite{bao2012finite}, the Crouzeix--Raviart method \cite{brenner2015forty},
the Taylor--Hood method \cite{ref:BoffiTaylorHood3triangles},
and the Scott--Vogelius method \cite{neilan2020stokes,lrsBIBih}.
The latter has the advantage that is it exactly divergence free, a property
that has been shown to be crucial for certain applications
\cite{christon2010consistency,gerbeau1997spurious}.

The results showcase a fundamental flaw with the Scott--Vogelius method when boundary conditions are imposed strongly on polygonally-approximated curved boundaries. This sheds new light on recent results with respect to grad-div stabilization of Taylor-Hood elements, which have been found to impact drag computation \cite{batugedara2022note}. As has been noted before in e.g. \cite{buffa2011isogeometric}, the combination of a divergence-free function with Dirichlet boundary conditions erroneously over-constrains the solution. We give, via a straightforward geometric argument, an example showing that this setup constrains the strain to be zero at the vertices of the mesh. This causes the strain to oscillate on the cylinder boundary, yielding spurious drag predictions.

To resolve this issue, we impose boundary conditions weakly using Nitsche's
method \cite{lrsBIBiy,ref:StenbergNitscheLagrange}. Numerical testing shows that this restores high accuracy. With this issue corrected, we find that the Scott--Vogelius--Nitsche method and Taylor--Hood method
agree to striking accuracy, and both of them have strong internal agreement
based on the verification metric studied here.
By contrast, P2-P0 and CR-P0 have significantly lower accuracy, even though
the verification metric indicates internal consistency.
Thus the verification metric can indicate failure of simulations methods,
but it cannot be used to guarantee high accuracy of such methods.
Otherwise said, it cannot be used as an error indicator
\cite{babuvska1979adaptive,ghesmati2019residual}.

With respect to validation, we discuss a well-known analytic solution for drag (Lamb's model) and discuss how the size of the domain influences comparisons of analytical, experimental and numerical predictions. To be more precise, we compare Lamb's model for cylinder drag at low
Reynolds numbers  \cite{lamb1993hydrodynamics} with drag computed via finite-element simulations.
We comment on an observed \cite{ref:cyldraglowRe} discrepancy
in this model and explain why it may have occurred.
We show that the size of the computational domain needs to be
quite large to match Lamb's model while still having good
agreement with earlier experimental results.
This supports observations that boundary effects in the experiments
lead to the observed discrepancy in \cite{ref:cyldraglowRe}.
For a discussion of both experimental and theoretical approaches to
determining drag on a cylinder, see \cite{ref:NigelGoldenfeldcylinderflow}.

\begin{figure}
\centerline{\includegraphics[width=5.0in,angle=0]{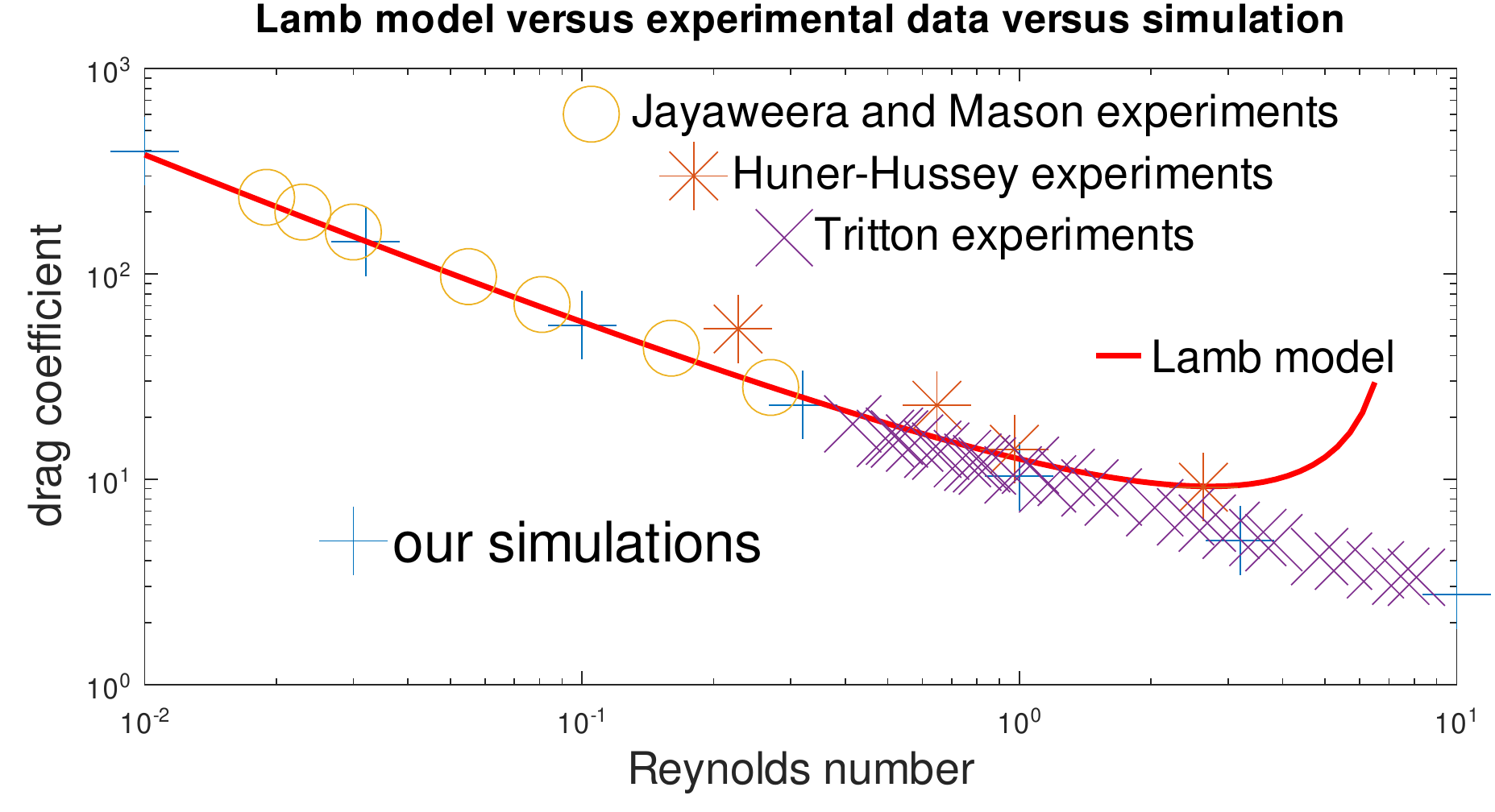}}
\vspace{-2mm}
\caption{Verification and validation performed by comparing analytical solutions (the Lamb model \eqref{eqn:lambodel}), numerical computations and experimental measurements of drag.
$\bigcirc$'s from \cite{ref:fallingcylinderdrag},
$\times$'s from \cite{tritton_1959}).
Simulation data is for a domain \eqref{eqn:boxcirl} with
$\ell_1=2400$, $\ell_2=7200$, $W=2400$, with $N_\Omega=128$ and $S=1024$.
}
\label{fig:potential}
\end{figure}

Computational drag verification and validation has broad applications.
It can test simulation software used for predicting boundary tangential stress,
a.k.a.~wall shear stress, which is of significant interest as a predictor
of cardiovascular disease
\cite{hung2015association,
reneman2006wall,
shaaban2000wall,
takehara2020abnormal,
von2016evaluation,
wallace1989software}.
Although the focus here is on low Reynolds numbers and Lamb's model,
the same numerical techniques have been applied to higher Reynolds number flows
in \cite{lrsBIBki}. As we showcase in this work, it can be challenging to compute boundary stress reliably using finite-element for domains with curved boundaries.
Computing cylinder drag thus provides an avenue for verifying simulations in a
broader context.

Finally, we note that the verification method studied here is applicable to any
simulation method, not just finite elements.
It does require that approximate solutions be represented as functions,
as is done in finite volume schemes, spectral methods, Chebyshev methods, and
Fourier methods, as well as finite element schemes.
It is also possible to relate finite difference and other
approximations to functions via interpolation.

The paper is organized in sections as follows.
Section \ref{sec:probset} recalls the Navier--Stokes equations and
boundary conditions of interest for flow past an obstacle.
It also recalls the definition of drag based on boundary integrals of
the stress and pressure.
Section \ref{sec:varformnav} discusses the verification approach
using an alternate way to compute drag.
It presents drag results for two different schemes, which agree to a remarkable degree.
Section \ref{sec:femapprox} describes the four finite element schemes
examined here.
Section \ref{sec:polybac} describes in detail the polygonal approximation
of a smooth boundary and gives an example that shows the pitfall that occurs
when using methods that are exactly divergence free.
Section \ref{sec:valamb} describes Lamb's model for drag at low Reynolds
numbers, relates this to the Stokes Paradox, and addresses a data
discrepancy in related experiments.

\section{Problem setting and model equations}
\label{sec:probset}

Suppose that $(\uu,p)$ is a solution of the stationary Navier--Stokes equations in a
domain $\Omega\subset\Reyuls^d$ containing an obstacle with boundary $\Gamma\subset\bo$:
\begin{equation} \label{eqn:firstnavst}
\begin{split}
-\nu\Delta \uu & + \uu\cdot\nabla\uu+\nabla p = \bfz\;\hbox{in}\;\Omega,\\
&\sdiv\uu =0\;\hbox{in}\;\Omega,
\end{split}
\end{equation}
together with boundary conditions
\begin{equation} \label{eqn:bceesnavst}
\uu=\gbc\;\hbox{on}\;\partial\Omega\backslash\Gamma,\quad
\uu=\bfz\;\hbox{on}\;\Gamma,
\end{equation}
in a domain $\Omega$, where $\nu$ is the kinematic viscosity.

The flow of fluid around an obstacle generates a force called drag, which is a fundamental concept in fluid dynamics \cite{lamb1993hydrodynamics}. It plays a critical role in determining the behavior of objects in flight and has been studied since the time of d'Alembert \cite{lrsBIBjn}. Drag is composed of two components: pressure drag $\beta_{p}$ and viscous drag $\beta_{v}$, which can be calculated by evaluating the following functions using $\vv=(1,0)$:
\begin{equation} 
\label{eq:firstdrag}
\begin{split}
\beta_v(\vv) &=\oint_{\bog} \big((\nu\du)\vv\big)\cdot\nn \,ds, \qquad
\beta_p(\vv)  =\oint_{\bog}  -p\vv\cdot\nn \,ds.
\end{split}
\end{equation}

\begin{figure}
\begin{subfigure}{0.9\textwidth}
\includegraphics[width=1.09\textwidth,angle=0]{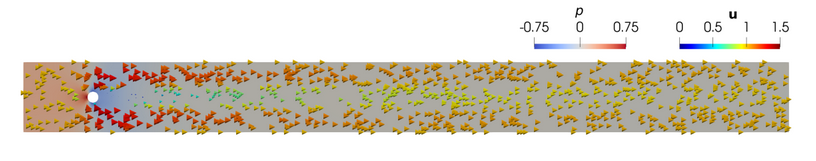}
\caption{}
\end{subfigure}
\begin{subfigure}{0.5\textwidth}
\includegraphics[width=0.99\textwidth,angle=0]{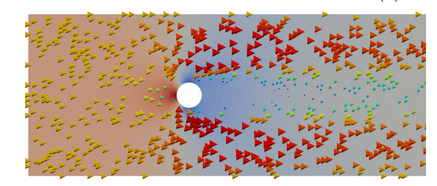}
\caption{}
\end{subfigure}
\begin{subfigure}{0.21\textwidth}
\includegraphics[width=1.15\textwidth,angle=0]{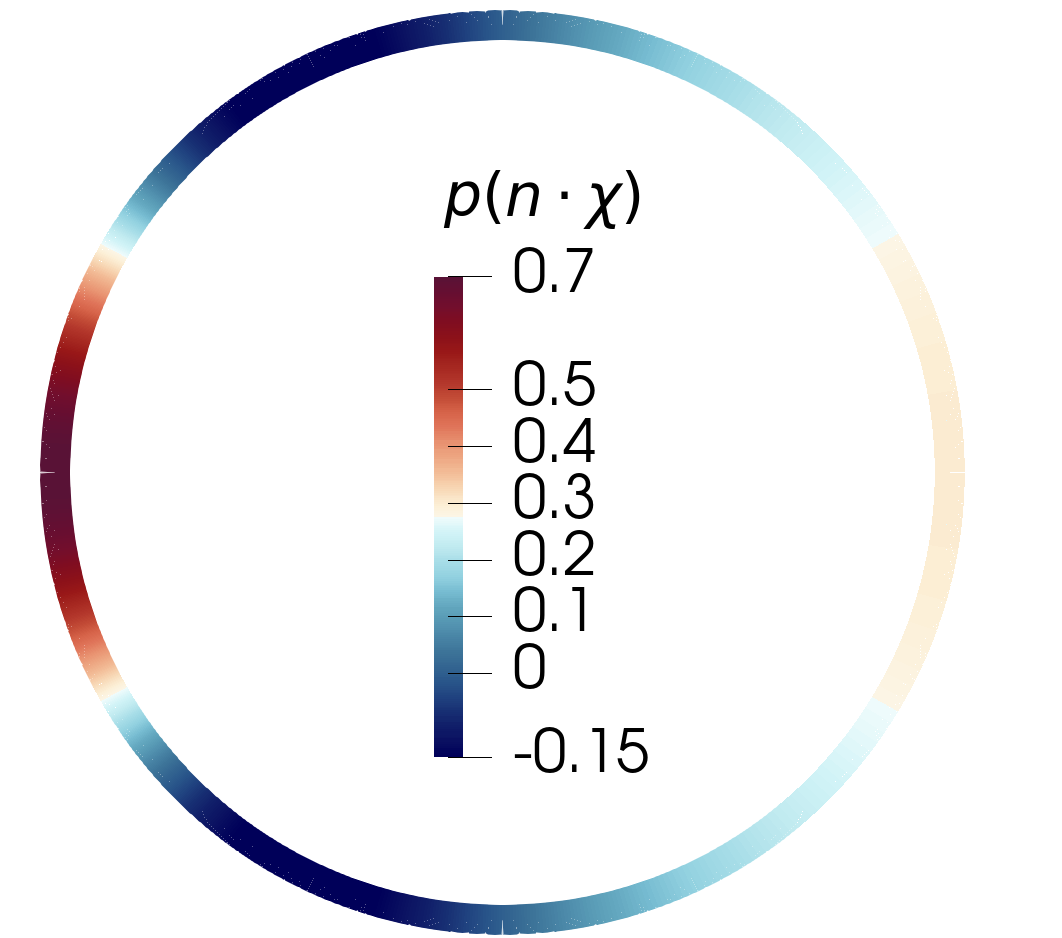}
\caption{}
\end{subfigure}
\begin{subfigure}{0.205\textwidth}
\quad\includegraphics[width=1.05\textwidth,angle=0]{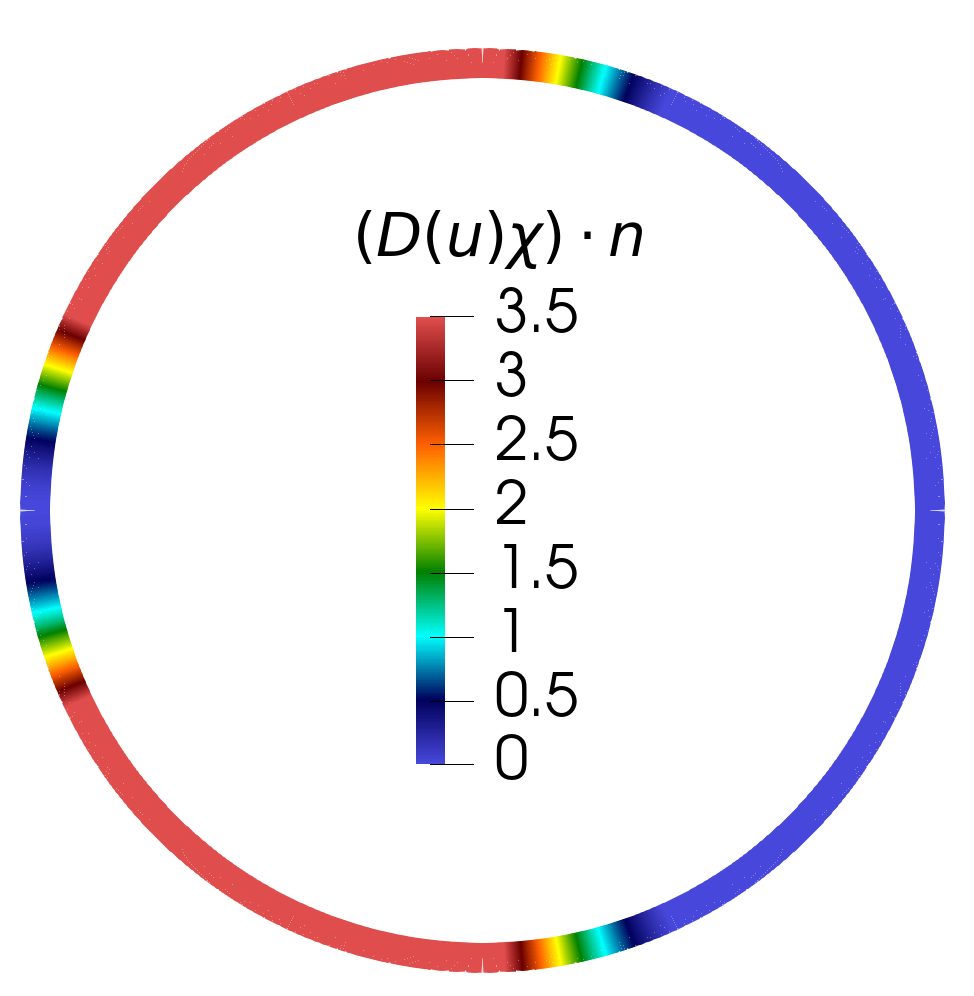}
\caption{}
\end{subfigure}
\caption{Steady flow around a cylinder at Reynolds number 50:
(a) full domain, (b) zoom around cylinder,
(c) pressure on cylinder boundary, (d) strain on cylinder boundary.
Computed with Taylor--Hood.}
\label{fig:flow50}
\end{figure}

A typical steady solution at Reynolds number 50 for flow around a cylinder is indicated in Figure \ref{fig:flow50}. The simulations were performed on the domain
\begin{equation} \label{eqn:oneomega}
\Omega=\set{(x,y)}{-12.8<x<128,\;|y|<W,\quad x^2+y^2>1}.
\end{equation}
As can be seen, the flow has to divert around the cylinder, causing shear strain at the top and bottom of the cylinder. This gives rise to the viscous drag. Moreover, a wake will form behind the cylinder, leading to heightened pressure at the front and decreased pressure at the back. This gives rise to the pressure drag.

Both types of drag can be challenging to approximate using,
e.g., finite elements.
In the next section, we discuss a validation method that can be used to evaluate numerical results.

\section{Verification: Two ways to compute drag}
\label{sec:varformnav}

In this section we show that for steady flows there are two equivalent ways of computing
drag, using either volumetric or surface integration.
Comparing the two computations provides a useful sanity check.
These methods are valid for any domains where the divergence theorem
\eqref{eqn:divtheoypa} is valid.
The connection between these methods is established through an
integration-by-parts formula \cite{john1997parallele}.
We start in Section \ref{sec:int-by-parts} by recalling this formula. With this in hand, we then define in Section \ref{sec:dragcompk} an internal inconsistency parameter that can be used as a necessary (but not sufficient) criterion for drag evaluation.

\subsection{Integration by parts}
\label{sec:int-by-parts}

Lemma 1.1 in \cite{lrsBIBiy} has an omission in the stated conditions, in that
the lemma applies only to divergence-free functions $\uu$, something that is clear
from the context but not explicitly stated.
Here we correct this by giving the result for general $\uu$.

\begin{lemma} \label{lem:stokinbypa}
Suppose that $\uu\in H^2(\Omega)^d$, $p\in H^1(\Omega)$, and $\vv\in H^1(\Omega)^d$.
Then
\begin{equation} \label{eqn:stokinbypa}
\begin{split}
\int_\Omega (-\nu\Delta\uu +\nabla p)\cdot\vv \,d\xx &=
\int_\Omega \frac{\nu}{2}\du:\dv + \nu\vv\cdot\nabla(\sdiv\uu) - p\sdiv\vv \,d\xx \\
&\qquad-\oint_{\bo} \big((\nu\du -pI)\vv\big)\cdot\nn \,ds,
\end{split}
\end{equation}
where $\nn$ is the outward normal to $\bo$.
\end{lemma}

Since the proof of Lemma \ref{lem:stokinbypa} is standard, we
relegate it to Appendix \ref{sec:proofstok}.
From \eqref{eqn:stokinbypa}, we obtain the following:

\begin{theorem} \label{thm:stokinbypa}
Suppose that $\uu\in H^2(\Omega)^d$, $p\in H^1(\Omega)$
is a solution of the steady Navier--Stokes equations
\begin{equation} \label{eqn:navstokipa}
-\nu\Delta\uu +\uu\cdot\nabla\uu+\nabla p =\bfz,\qquad \sdiv\uu=0
\end{equation}
in $\Omega$.
Define linear functionals
\begin{equation} \label{eqn:linfuncipa}
\begin{split}
\omega(\vv)&=\int_\Omega \frac{\nu}{2}\du:\dv +(\uu\cdot\nabla\uu)\cdot\vv-p\sdiv\vv\,d\xx,\\
\beta(\vv) &=\oint_{\bo} \big((\nu\du -pI)\vv\big)\cdot\nn \,ds,
\end{split}
\end{equation}
Then $\omega(\vv)=\beta(\vv)$ for all $\vv\in H^1(\Omega)^d$.
\end{theorem}

The functionals $\omega$ and $\beta$ are named according to the integration domains they involve, either the domain $\Omega$ or the boundary $\beta$. The relation $\omega(\vv)=\beta(\vv)$ holds for all $\vv\in H^1(\Omega)^d$, providing a convenient sanity check for the accuracy of computations.

\begin{proof}
The boundary term in \eqref{eqn:stokinbypa} is equivalently
$$
\big((\nu\du -pI)\vv\big)\cdot\nn=\nn^t\big(\nu\du -pI\big)\vv.
$$

We apply \eqref{eqn:stokinbypa} to get
\begin{equation} \label{eqn:prffuncipa}
0=\int_\Omega (-\nu\Delta\uu+\uu\cdot\nabla\uu+\nabla p)\cdot\vv\,d\xx
=\omega(\vv) -\beta(\vv) ,
\end{equation}
where we used the fact that $\sdiv\uu=0$.
Note that if $\sdiv\vv=0$, then
$$
\omega(\vv)=\int_\Omega \frac{\nu}{2}\du:\dv +(\uu\cdot\nabla\uu)\cdot\vv\,d\xx.
$$
\end{proof}

\subsection{Drag computation by volumetric or surface integration}
\label{sec:dragcompk}

The formula \eqref{eq:firstdrag} computes the drag via the function $\beta(\vv)$
with $\vv=(1,0)$, which involves integration over the surface $\Gamma$.
Via the integration by parts formula introduced in the last section, we will now construct an equivalent formulation using volume integration. To achieve this we need to suitably extend $\vv=(1,0)$ to the whole domain.

Drag can be defined, e.g., for the domain $\Omega$ defined
in \eqref{eqn:boxcirl}, as $\beta(\bchi)$, where $\bchi=(1,0)$
on the cylinder and zero on the outer rectangular boundary.
This is also the correct definition of drag for a time-dependent flow.
We let $\bchi$ be a solution of the Stokes
equations with the required boundary conditions.

In practice, we approximate $\bchi$ via a finite-element function $\bchi_h$
using the same scheme that we use to approximate \eqref{eqn:firstnavst}.
Similarly we define $\beta_h$ and $\omega_h$ using $\uu_h$
instead of $\uu$ and $p_h$ instead of $p$.
Then we take $\beta_h(\bchi_h)$ as our approximation to the drag
coefficient.
We have $\sdiv\bchi=0$ for the Scott--Vogelius method, which
allows us to drop the pressure term in $\omega$.
Of course, $\beta_h(\bchi_h)-\omega_h(\bchi_h)$ will be small,
but not zero.

Error estimates in terms of Sobolev norms have been given in \cite{ref:errestdragtimedep},
but not including the errors associated with boundary approximation.
The error using $\omega$ can be substantially less because the integrand
in \eqref{eqn:linfuncipa} is close to zero near $\Gamma$, so there
is little error due to integrating over just $\Omega_h$.
Since the evaluation of $\omega$ only involves the domain, it bypasses certain issues related to boundary discretization and boundary conditions. Indeed, as noted in \cite{von2016evaluation}, $\omega$ tends to yield more accurate drag values. A mismatch between $\omega$ and $\beta$ typically indicates issues related to
computations near the boundary.

In addition to \eqref{eq:firstdrag}, define linear functionals
\begin{equation} \label{eqn:sublinfunc}
\begin{split}
\omega_v(\vv)&=\int_\Omega \frac{\nu}{2}\du:\dv \,d\xx,\qquad
\omega_a(\vv) =\int_\Omega (\uu\cdot\nabla\uu)\cdot\vv-p\sdiv\vv\,d\xx.
\end{split}
\end{equation}
We will omit the subscript $h$ for the discrete approximations.

Let $\epsilon$ denote the \textit{internal inconsistency measure}
\begin{align}
\epsilon = \frac{\omega-\beta}{\omega} \label{eq:incon}
\end{align}
From this we can define a straightforward validation method, which involves checking that $\epsilon \rightarrow 0$ as the mesh size decreases.  As we will see in Table \ref{tabl:badmat}, this is necessary but not sufficient criterion
for accurate drag computation.

\section{Finite element approximation}
\label{sec:femapprox}

In this section, we measure the internal inconsistency \eqref{eq:incon} for different finite element approximations of the drag. To start, we give here the finite element approximation of the Navier-Stokes equations. In Section \ref{sec:dragcompk}, we show the inconsistency parameter for Taylor-Hood, Scott-Vogelius,
P2-P0, and Crouzeix--Raviart
finite elements. As we will see, the internal inconsistency varies substantially between finite element families.

The Navier-Stokes equations can be written in a weak, or variational, form as follows:

Find $\vv \in (H^1(\Omega))^d$ and $p \in L^2(\Omega)$ such that

\begin{align*}
a(\vv, \ww) + b(\ww, p) &= 0 &\quad &\forall \ww \in (H^1(\Omega))^d, \\
b(\vv, q) &= 0 &\quad &\forall q \in L^2(\Omega),
\end{align*}
where the bilinear forms $a(\cdot, \cdot)$ and $b(\cdot, \cdot)$ are defined as:
\begin{align*}
a(\vv, \ww) &= \int_\Omega \frac{\nu}{2} \mathcal{D}(\nabla \vv) : \mathcal{D}(\nabla \ww)  \,d\xx, \\
b(\vv, q) &= -\int_\Omega q (\nabla \cdot \vv) \,d\xx,
\end{align*}
where $\dv=\nabla\vv+\nabla\vv^t$.

Finite element approximations are obtained by selecting suitable discretizations $V_h$ and $Q_h$ of $(H^1(\Omega))^d$ and $L^2(\Omega)$. In this work we consider four common choices, namely
\begin{itemize}
\item the Taylor--Hood element, which is a combination of piecewise quadratic velocity and piecewise linear pressure functions, 
\item the P2-P0 element, which is a combination of piecewise quadratic velocity and piecewise constant pressures, 
\item the Crouzeix--Raviart element (CR-P0), which is a combination of linear,
non-conforming velocities and piecewise constant pressures,
\item and the Scott--Vogelius element, which is uses Lagrange elements of degree 4 for
the velocity and the divergence of this space for the pressure space.
\end{itemize}
For further details around the finite elements we refer to the code repository \cite{code}.

As we will see, obtaining good finite element approximations of the drag \eqref{eq:firstdrag} can be challenging in practice.
If the curved boundary is approximated by polygons, then the derivatives
on these polygons can be substantially in error.
In scalar elliptic problems, it is known that the mean-squared error,
averaged over the domain, of the derivatives is order $h^{3/2}$ for a mesh of size $h$ \cite{lrsBIBaa}.
Thus it is reasonable to expect that the error on the boundary for the
derivatives would go to zero like a positive power of $h$.
In the next section, we will use the inconsistency parameter $\epsilon$ to show
that this is obtained with suitable methods.
But in Table \ref{tabl:mismat}, we show that this is not the case for
vector-valued approximations satisfying an incompressibility constraint.


\begin{table}\footnotesize
\begin{center}
\begin{subtable}{0.45\textwidth}
\centering
\begin{tabular}{|r|c|c|c|}\hline
$R$ &  $N_\Omega$ & $\omega$ & $10^2\epsilon$  \\
\hline
1& 32& 26.195 &  0.037 \\
1& 64& 26.101 &  0.015 \\
1& 128& 26.078 &  0.007 \\ \hline
10&  32& 4.117 &  0.036 \\
10&  64& 4.104 &  0.014 \\
10& 128& 4.099 &  0.005 \\ \hline
50&  32& 1.880 &  0.035 \\
50&  64& 1.873 &  0.013 \\
50& 128& 1.868 &  0.004 \\ \hline
100&  32& 1.489 &  0.040 \\
100&  64& 1.486 &  0.013 \\
100&  128& 1.471 &  0.004 \\
\hline
\end{tabular}
\caption{TH elements with Dirichlet boundary condition on $\Gamma_h$}
\end{subtable}
\hspace{2em}
\begin{subtable}{0.45\textwidth}
\begin{tabular}{|r|c|c|c|c|}\hline
$R$ & $N_\Omega$ & $\omega$ & $10^2\epsilon$ \\
\hline
  1  & 32&  25.016 & -2.3\%   \\
  1  & 64& 26.099 & .018\%    \\
  1  & 128& 26.079 & .0084\%  \\
\hline
  10 &  32& 3.854  & -5.6\%  \\
  10 &  64&  4.103  & .017\%  \\
  10 &  128& 4.099  & .0075\%  \\
\hline
  50 &  32&  1.737  & -8.1\%  \\
  50 &  64& 1.874  & .015\%   \\
  50 &  128& 1.868  & .0055\% \\
\hline
  100 &  32& 1.386  & -7.6\%   \\
  100 &  64& 1.487  & .016\%     \\
  100 &  128& 1.472  & .0052\%   \\
\hline
\end{tabular}
\caption{SV elements, Nitsche boundary condition on $\Gamma_h$ with $\mu=10^6$}
\end{subtable}
\caption{Successful consistency checks for drag computation using Scott-Vogelius (SV) and Taylor-Hood elements (TH) elements. The drag approximation converges to the same value for both elements, and the inconsistency parameter $\epsilon$ decreases with mesh refinement. Inconsistency was computed using functionals \eqref{eqn:linfuncipa} and \eqref{eqn:sublinfunc} applied to $\bchi$ for various mesh size parameters $N_\Omega$ and  Reynolds numbers $R$ for a channel half-width $W=6.4$, on the domain in \eqref{eqn:oneomega}, with inconsistency parameter $\epsilon=(\omega-\beta)/\omega$. }
\label{tabl:goodmat}
\end{center}
\end{table}

\begin{table}
\begin{center}
\begin{center}
\begin{tabular}{|r|c|c|c|c|c|c|c|c|}\hline
$R$ & $N_\Omega$ &$\beta$ & $\beta_p$ & $\beta_v$ & $\omega$ & $\omega_p$ & $\omega_v$ & $10^2 \epsilon$
\\
\hline
1 & 34& 16.8463 & 13.6859 & 3.16 & 26.1143 & 0.4738 & 25.64  & 34.29\% \\
1 & 64& 16.6780 & 13.7184 & 2.96 & 26.0960 & 0.4736 & 25.62  & 34.90 \%\\
1 & 128& 16.7128 & 13.6472 & 3.07 & 26.0850 & 0.4735 & 25.61  & 34.74 \%\\
\hline
10 & 34& 2.8034 & 2.3701 & 0.433  &  4.1029 & 1.5389 & 2.564 & -9.34 \% \\
10 & 64&  2.7866 & 2.3732 & 0.413   &  4.1002 & 1.5380 & 2.562 & -8.77\% \\
10 & 128& 2.8094 & 2.3644 & 0.445  &  4.0985 & 1.5374 & 2.561  & -9.69\% \\
\hline
50 & 34& 1.4303 & 1.2880 & 0.142  &  1.8683 & 1.3555 & 0.5128 & -178 \% \\
50 & 64& 1.4281 & 1.2898 & 0.138  &  1.8676 & 1.3552 & 0.5124 & -178 \% \\
50 & 128& 1.4277 & 1.2866 & 0.141  &  1.8688 & 1.3546 & 0.5122 & -178 \%\\
\hline
100& 34& 1.1911 & 1.1012 & 0.0899  &  1.4717 & 1.2153 & 0.2564 & -364\% \\
100& 64& 1.1918 & 1.1036 & 0.0882  &  1.4720 & 1.2157 & 0.2562  & -365\% \\
100& 128&  1.1953 & 1.1013 & 0.0940  &  1.4712 & 1.2150 & 0.2561 & -662\% \\
\hline
\end{tabular}
\end{center}
\vspace{-1mm}
\caption{Consistency checks for drag computation using Scott-Vogelius elements with
strongly imposed homogeneous Dirichlet conditions on $\Gamma_h$.
The results
are strongly inconsistent, indicating numerical issues;
we explain these results in Section \ref{sec:polybac}.
The cylinder boundary mesh was heavily refined, with $N_\Gamma=1000 N_\Omega$.}
\label{tabl:mismat}
\end{center}
\end{table}

\begin{table}
\begin{center}
\begin{subtable}{0.9\textwidth}\footnotesize
\begin{center}
\begin{tabular}{|r|c|c|c|c|c|c|c|c|}\hline
$R$ & $N$ & $\beta_p$ & $\beta_v$ & $\beta$ & $\omega$ & $10^2\epsilon$   \\
\hline
\multirow{4}{*}{1} & 32& 13.255 \scriptsize(3.0\%) & 11.781 \scriptsize(5.3\%) & 25.006  & 24.473 &  -2.2\% \\
& 64& 13.390 \scriptsize(1.8\%) &  12.369 \scriptsize(0.6\%) & 25.760  & 25.764 &  .017\% \\
&128& 13.522 \scriptsize(0.8\%) &  12.423 \scriptsize(0.2\%) & 25.944  & 25.946 &  .0072\% \\
& 256& 13.587 \scriptsize(0.3\%) &  12.443 \scriptsize (0.0\%) & 26.031 & 26.031  &  0.003 \%  \\
\hline
\multirow{4}{*}{10}  & 32& 2.344 \scriptsize(0.9\%) &  1.677 \scriptsize(3.1\%) & 4.020  & 3.814 &  -5.4\%   \\
& 64& 2.359 \scriptsize(0.3\%) &  1.720 \scriptsize(0.6\%) & 4.079  & 4.080 &  .022\%   \\
&128& 2.385 \scriptsize(0.8\%) &  1.730 \scriptsize(0.0\%) & 4.115  & 4.115 &  -.009\%   \\
& 256& 2.377 \scriptsize(0.5\%) &  1.731 \scriptsize (0.0\%) & 4.108  & 4.108 &  -0.003 \\
\hline
\multirow{4}{*}{50}  & 32& 1.357 \scriptsize(5.4\%) &  0.576 \scriptsize(0.4\%) & 1.933  & 1.791  &  -7.9\% \\
& 64& 1.333 \scriptsize(3.5\%) &  0.576 \scriptsize(0.3\%) & 1.909  & 1.910  &  -.032\% \\
&128& 1.367 \scriptsize(6.1\%) &  0.587 \scriptsize(1.5\%) & 1.953  & 1.954  &  .013\% \\
&  256& 1.352\scriptsize (5.0\%) &  0.581 \scriptsize (0.5\%) & 1.933 & 1.933 &  0.006 \\
\hline
\multirow{4}{*}{100}  & 32& 1.181 \scriptsize(7.1\%) &  0.367 \scriptsize(1.1\%) & 1.548  & 1.450  &  -6.8\% \\
& 64& 1.168 \scriptsize(5.9\%) &  0.368 \scriptsize(0.0\%) & 1.536  & 1.537  &  .039\% \\
&128& 1.213 \scriptsize(10.0\%) &  0.377 \scriptsize(2.5\%) & 1.590 & 1.590 &  .016\%   \\
& 256& 1.201 \scriptsize(8.9\%) &  0.374 \scriptsize(1.6\%) & 1.575 & 1.575 &  0.008 \\
\hline
\end{tabular}
\end{center}
\vspace{-5mm}
\caption{P2-P0 elements with Dirichlet boundary conditions on $\Gamma_h$.}
\label{tabl:ptoopz}
\end{subtable}

\vspace{1em}
\begin{subtable}{0.9\textwidth}\footnotesize
\begin{center}\begin{tabular}{|r|c|c|c|c|c|c|c|c|}\hline
R & N & $\beta_p$ & $\beta_v$ & $\beta$ & $\omega$ & $10^2 \epsilon$ \\ \hline
\multirow{4}{*}{1}  & 32& 10.443 \scriptsize (23.4\%) &  11.195 \scriptsize(10.0\%) & 21.638  & 21.256  &  -1.8\% \\
& 64& 11.635 \scriptsize (14.6\%) &  11.767 \scriptsize(5.4\%) & 23.403 & 23.201 &  -0.87\% \\
& 128& 12.803 \scriptsize (6.1\%) &  12.083 \scriptsize(2.9\%) & 24.886  & 24.788 &  -0.39\% \\
& 256& 13.419 \scriptsize (1.6\%) &  12.182 \scriptsize(2.1\%) & 25.602 & 25.631  &  1.1e-01 \\  \hline
\multirow{4}{*}{10}& 32& 2.181 \scriptsize (7.8\%) &  1.606 \scriptsize (7.2\%) & 3.787 & 3.700 &  -2.4\% \\
& 64& 2.302 \scriptsize  (2.7\%) &  1.683 \scriptsize (2.8\%) & 3.985 & 3.925 &  -1.5\% \\
& 128& 2.373 \scriptsize (0.3\%) &  1.726 \scriptsize (0.3\%) & 4.100 & 4.073 &  -0.66\% \\
& 256 & 2.373 \scriptsize  (0.3\%) &  1.731 \scriptsize (0.0\%) & 4.104  & 4.093 &  -0.28 \\  \hline
\multirow{4}{*}{50}& 32& 1.241 \scriptsize (3.7\%) &  0.535 \scriptsize (7.5\%) & 1.776 & 1.756  &  -1.1\% \\
& 64& 1.304 \scriptsize (1.2\%) &  0.579 \scriptsize (0.1\%) & 1.883 & 1.848  &  -1.9 \%\\
&128& 1.295 \scriptsize (0.6\%) &  0.594 \scriptsize (2.8\%) & 1.889 & 1.875  &  -0.76\% \\
& 256& 1.308 \scriptsize (1.5\%) &  0.594 \scriptsize (2.8\%) & 1.902  & 1.895 &  -0.39 \\ \hline
\multirow{4}{*}{100}& 32& 1.062 \scriptsize (3.7\%) &  0.335 \scriptsize (9.0\%) & 1.397 & 1.373 &  -1.7\% \\
& 64& 1.126 \scriptsize (2.1\%) &  0.374 \scriptsize (1.5\%) & 1.500 & 1.469 &  -2.1\% \\
& 128& 1.101 \scriptsize (0.2\%) &  0.376 \scriptsize (2.2\%) & 1.477 & 1.465 &  -0.8\% \\
& 256 & 1.106 \scriptsize (0.3\%) &  0.387 \scriptsize (5.0\%) & 1.493 & 1.487 &  -0.36 \\ \hline
\end{tabular}
\end{center}
\caption{CR-P0 elements with Dirichlet boundary conditions on $\Gamma_h$.}
\label{tabl:crpeeo}
\end{subtable}

\end{center}
\vspace{-5mm}
\caption{Consistency checks for drag computation using P2-P0 and CR-P0 elements,
both with Dirichlet boundary conditions on $\Gamma_h$.
Both P2-P0 and CR-P0 yield internally consistent values, in the sense that
the inconsistency parameter $\epsilon$ decreases with mesh refinement.
Comparing against the drag values from Table \ref{tabl:goodmat}, however, shows that
these discretizations are converging to a \textit{different} value.
The percentage errors in parentheses for $\beta_p$, $\beta_v$, $\beta$, $\omega$ are based on a comparison with values computed using Taylor--Hood.
For P2-P0, the main error occurs in the pressure drag, while
for CR-P0, the main error occurs in the viscous drag.
This indicates that decreasing internal consistency is necessary but not sufficient to provide convergence.
 The consistency check was performed by comparing functionals \eqref{eqn:linfuncipa} and \eqref{eqn:sublinfunc} applied to $\bchi$ for various mesh size parameters $N_\Omega$ and  Reynolds numbers $R$ for a channel half-width $W=6.4$, on the domain
in \eqref{eqn:oneomega}, with inconsistency parameter  $\epsilon=(\omega-\beta)/\omega$.}
\label{tabl:badmat}
\end{table}

\subsection{Results for Taylor-Hood, Scott-Vogelius, P2-P0, and Crouzeix--Raviart}
\label{sec:applcompk}

Table \ref{tabl:goodmat} shows some results of computation of
functionals \eqref{eqn:linfuncipa} 
with $\uu_h$ and $p_h$ computed with two different methods
for a channel half-width $W=6.4$,
for various  Reynolds numbers $R$ and mesh size
parameters $N_\Omega$.
Both involve a polygonal approximation of the cylinder with a number of
segments $N_\Gamma$ equal to $8N_\Omega$.
For the Taylor--Hood method, homogeneous Dirichlet conditions are
imposed on the polygonal boundary.
For the Scott--Vogelius method, Nitsche's method is used to
enforce boundary conditions weakly on the cylinder.
In both cases, the various functionals are applied to
the approximate Stokes solution $\bchi_h$.
For a number of circle segments $N_\Gamma$ greater than $8N_\Omega$, the computed data agreed with
the data shown to the number of digits reported in Table \ref{tabl:goodmat}(a).

The close agreement of the Taylor--Hood and Scott--Vogelius--Nitsche methods,
together with the verification results using the approach in
section \ref{sec:dragcompk}, gives very high confidence in the correctness
of the implementation of both methods, both of which were implemented in
(different versions of) FEniCS \cite{FEniCSbook,lrsBIBih}.
Note that Taylor--Hood uses a piecewise-quadratic approximation
of the velocity, whereas Scott--Vogelius uses a piecewise-quartic approximation.
For this reason, computations with mesh size parameters $N_\Omega$ for Scott--Vogelius
were compared with mesh size parameters $2N_\Omega$ for Taylor--Hood, thus ensuring that the number of degrees of freedom were comparable.

\begin{figure}
\centerline{\includegraphics[width=1.6in,angle=0]{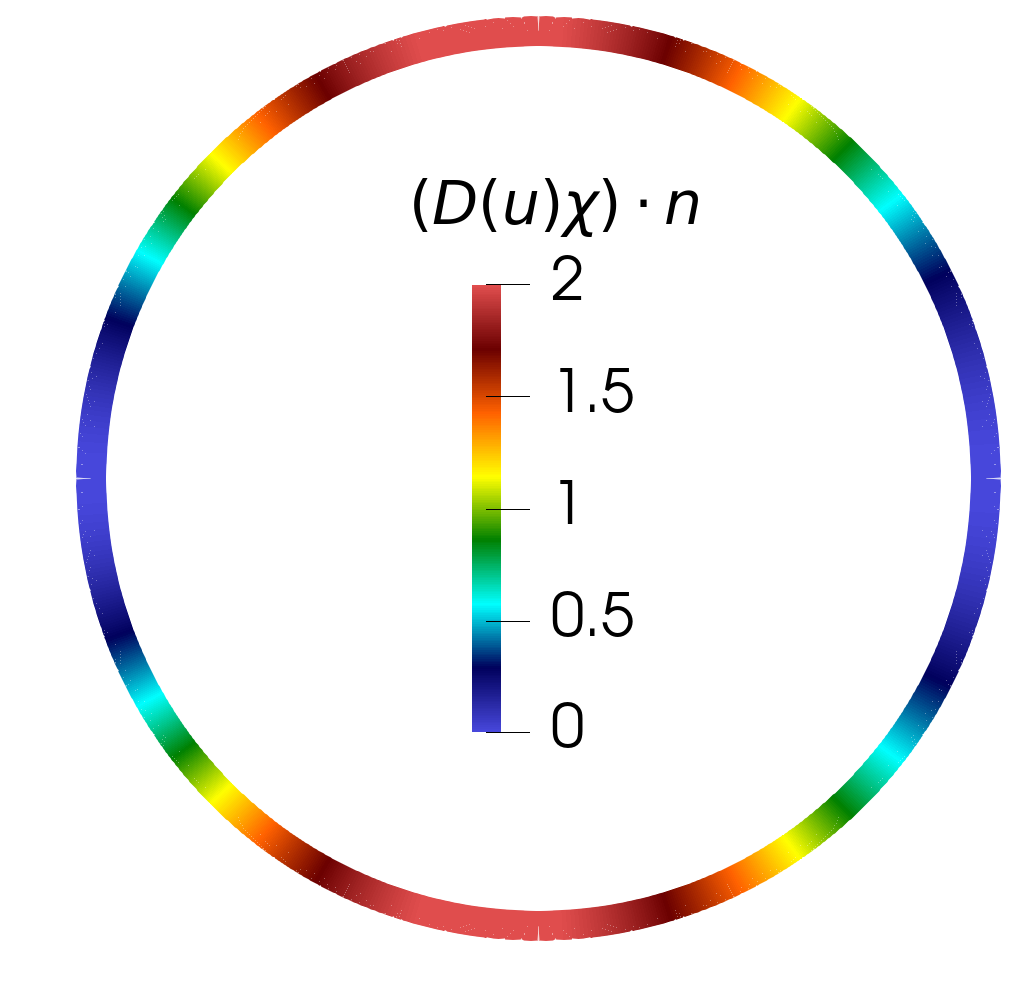}
           \includegraphics[width=1.6in,angle=0]{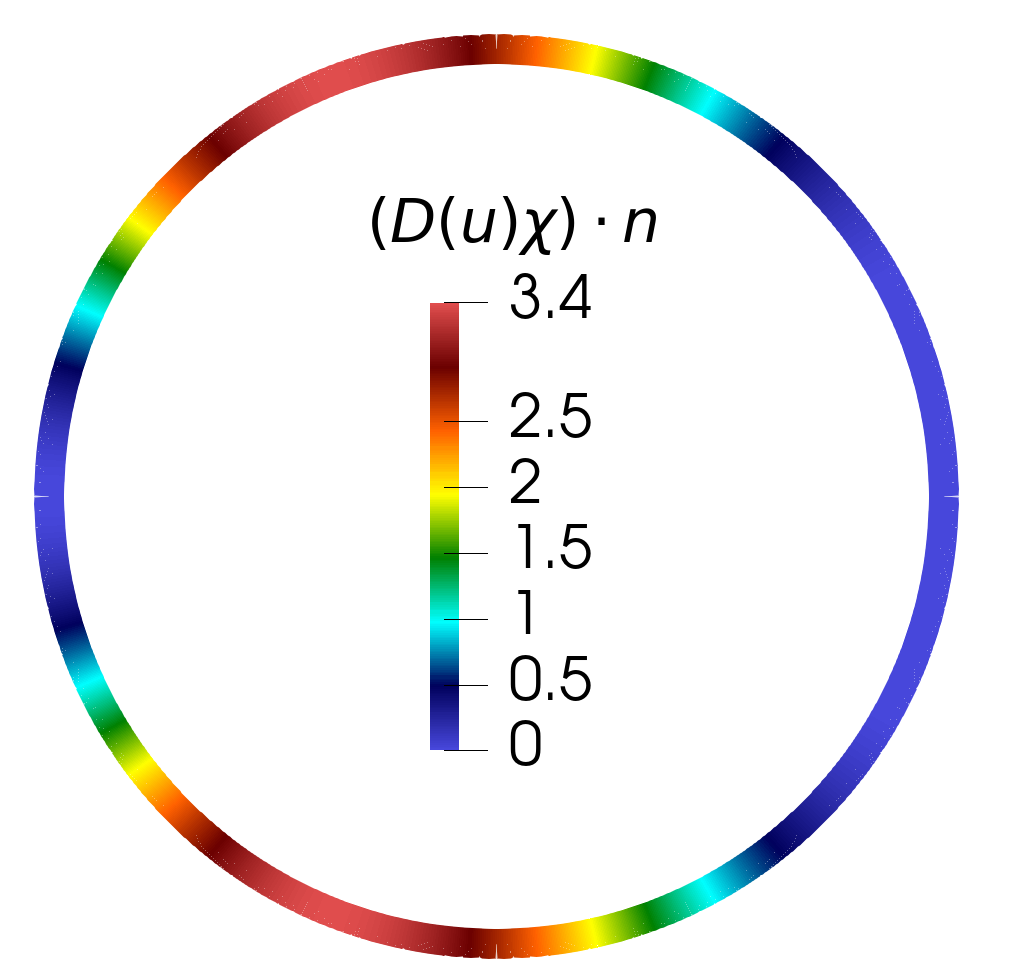}
           \includegraphics[width=1.55in,angle=0]{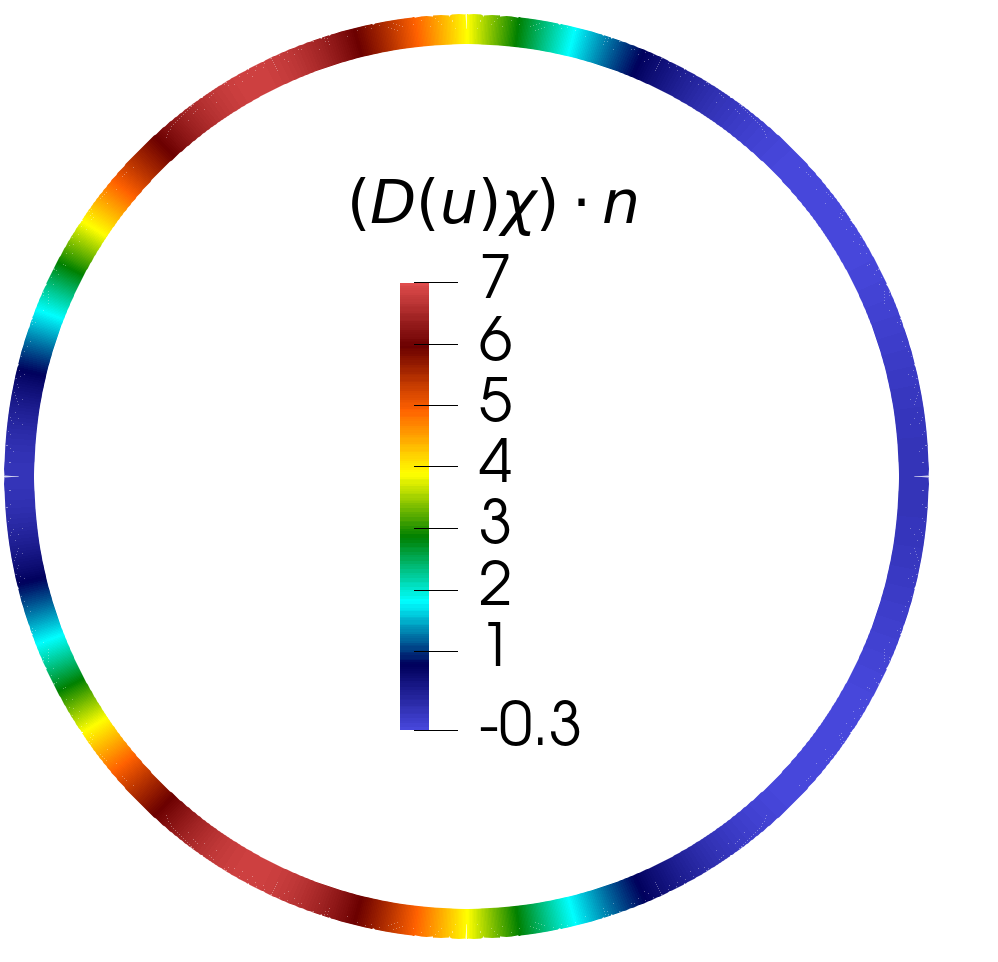}
           \includegraphics[width=1.5in,angle=0]{vdrag3_TH.png}}
\centerline{\includegraphics[width=1.5in,angle=0]{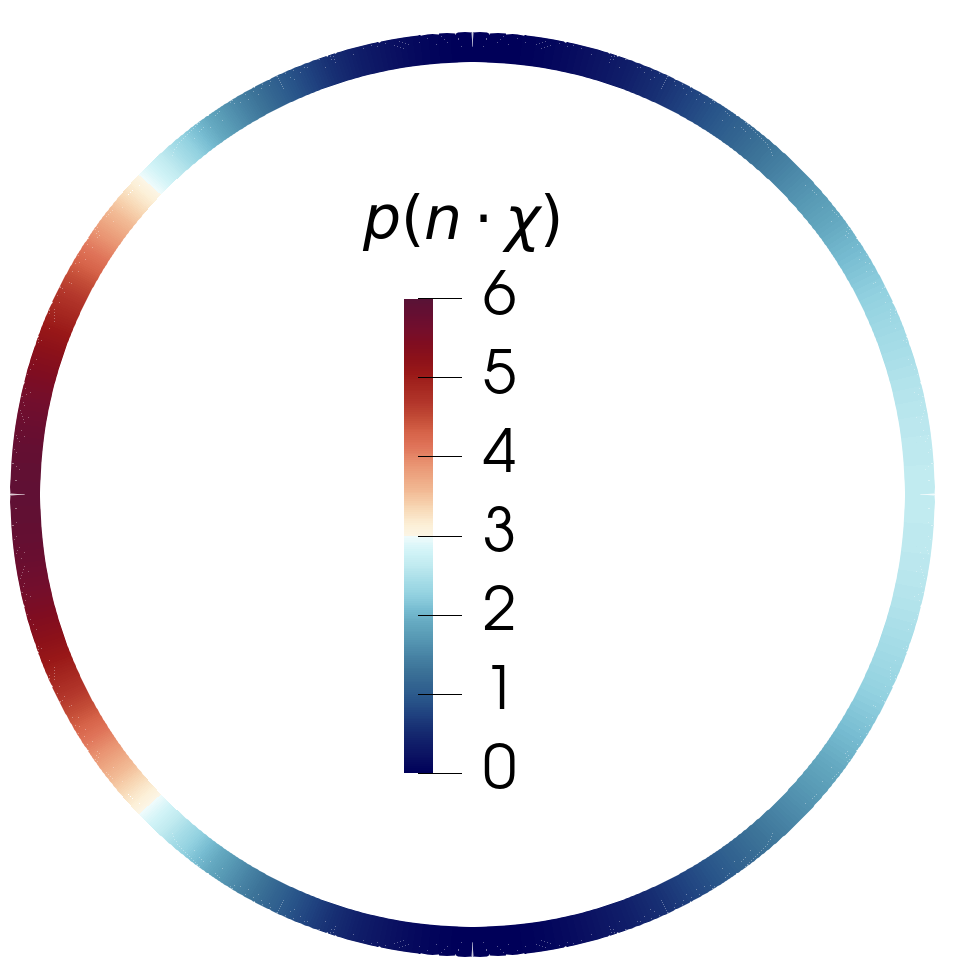}
           \includegraphics[width=1.65in,angle=0]{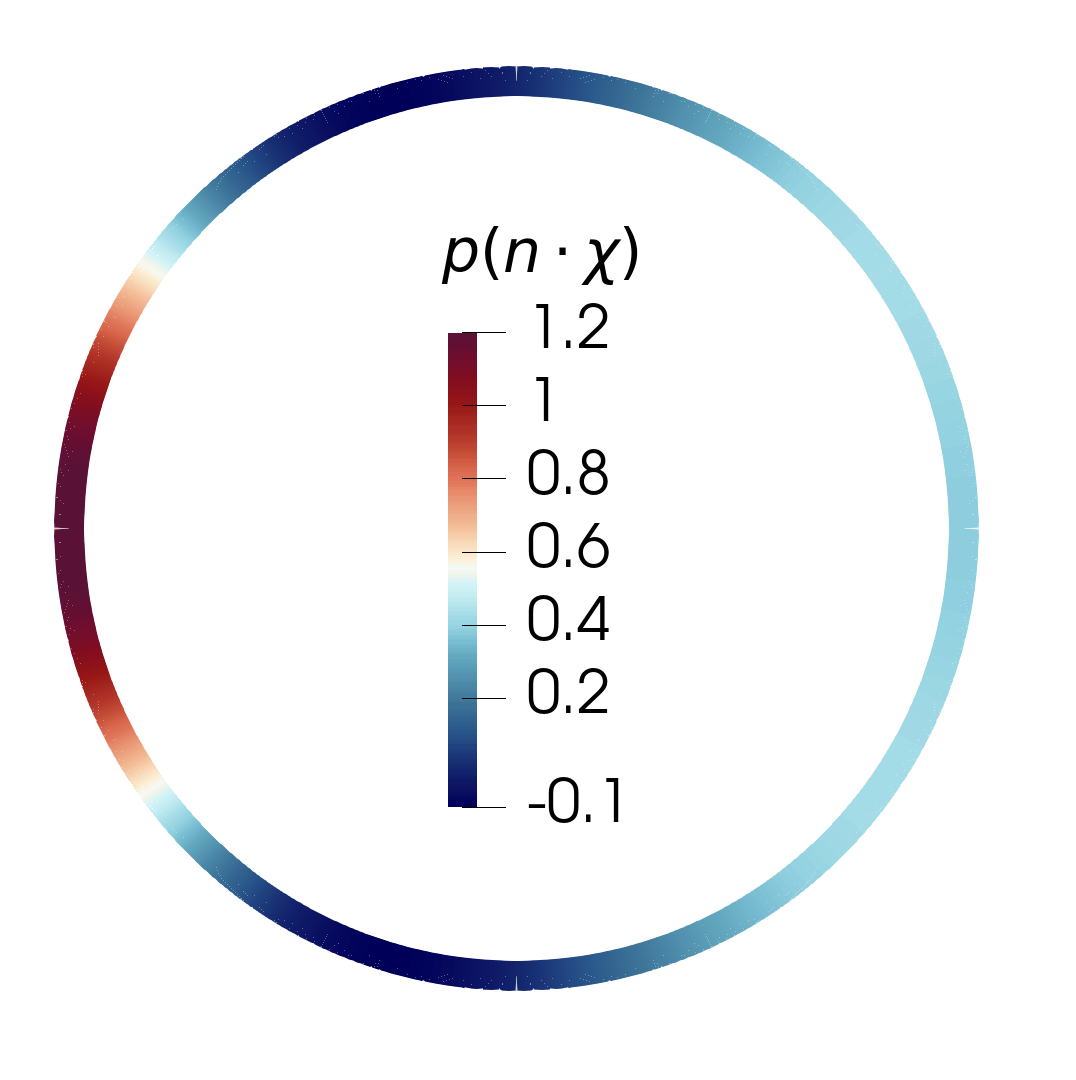}
           \includegraphics[width=1.6in,angle=0]{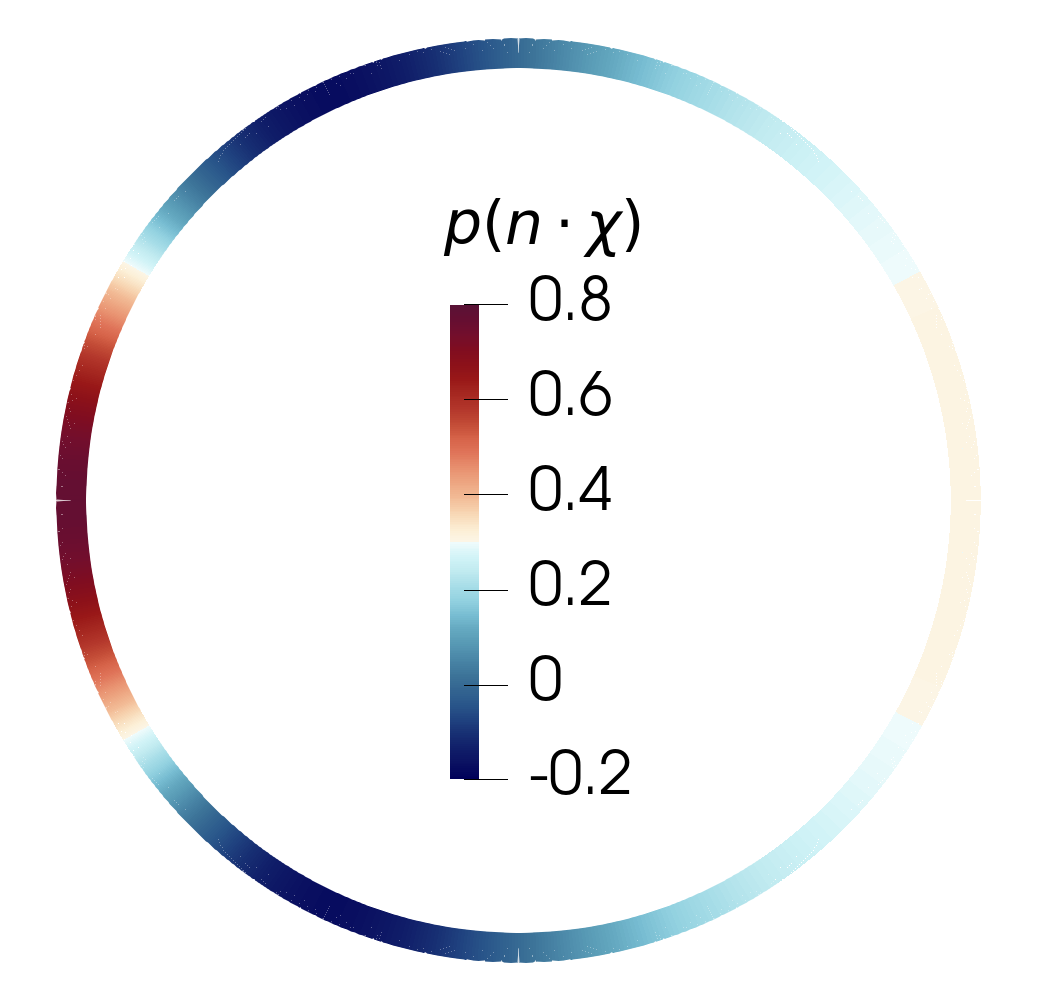}
           \includegraphics[width=1.6in,angle=0]{pdrag3_TH.png}}
\vspace{-2mm}
\caption{Cylinder strain (top) and pressure (bottom)
at $R=1,10,50,100$ (left to right).
Computed using Taylor--Hood.}
\label{fig:bdrag}
\end{figure}

The performance of the two methods is largely the same, with only small
differences.
For example, $\omega$ is computed accurately on coarse meshes by Scott--Vogelius,
whereas Taylor--Hood requires mesh refinement to reach comparable accuracy.
On the other hand, $\beta$ is less accurate for Scott--Vogelius on coarser
meshes, while Taylor--Hood achieves more accuracy for fine meshes.
A plot of the cylinder strain and pressure values computed with Taylor--Hood
is given in Figure \ref{fig:bdrag}.

\begin{figure}
\centerline{\includegraphics[width=1.7in,angle=0]{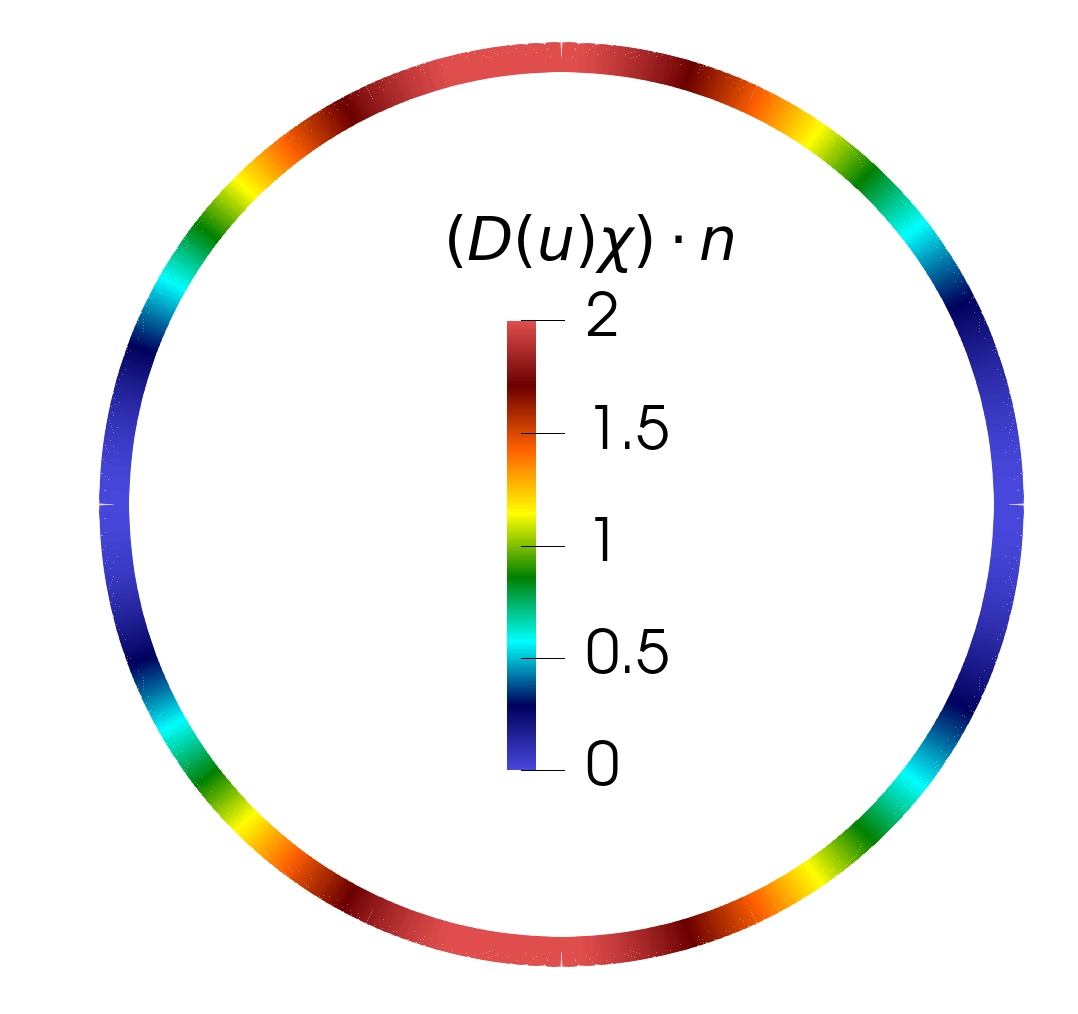}
           \includegraphics[width=1.5in,angle=0]{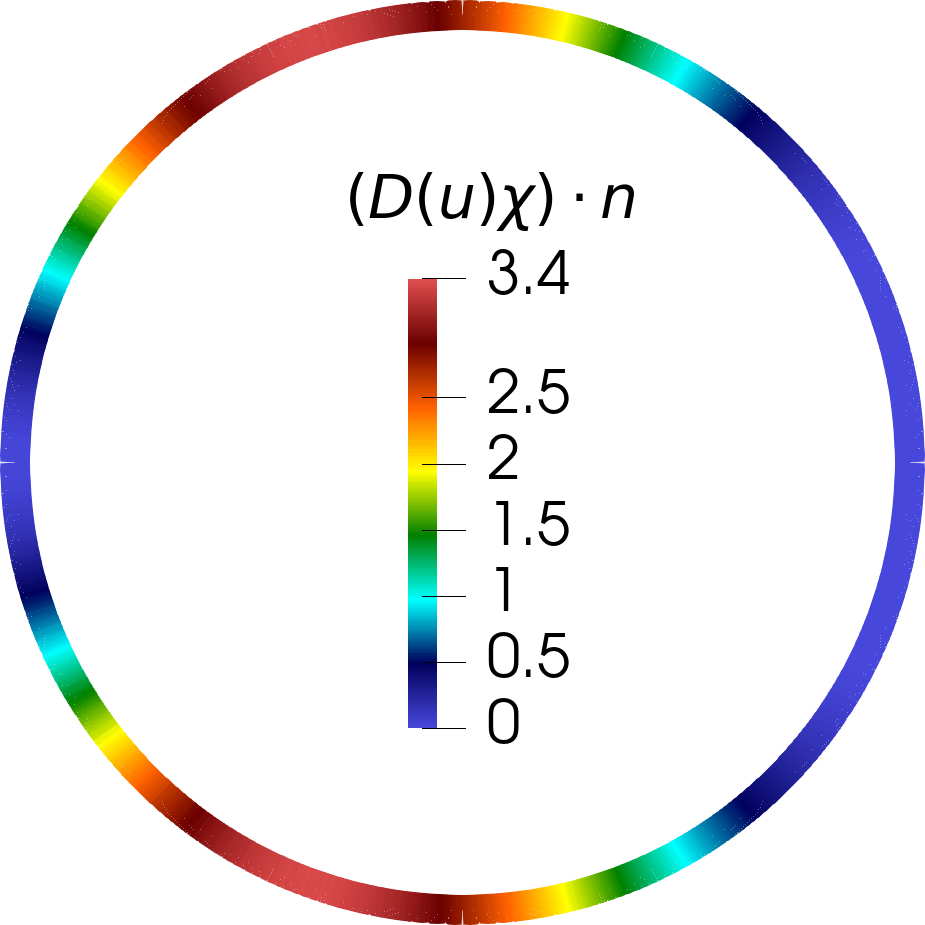}
           \includegraphics[width=1.5in,angle=0]{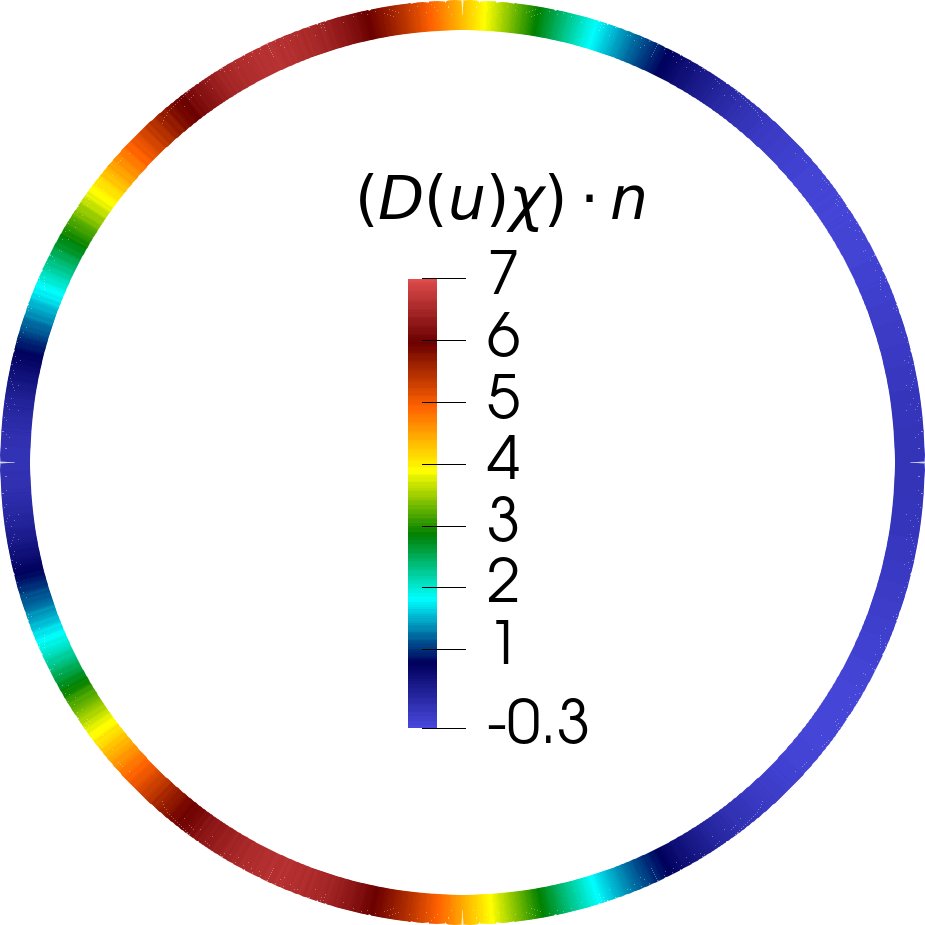}
           \includegraphics[width=1.5in,angle=0]{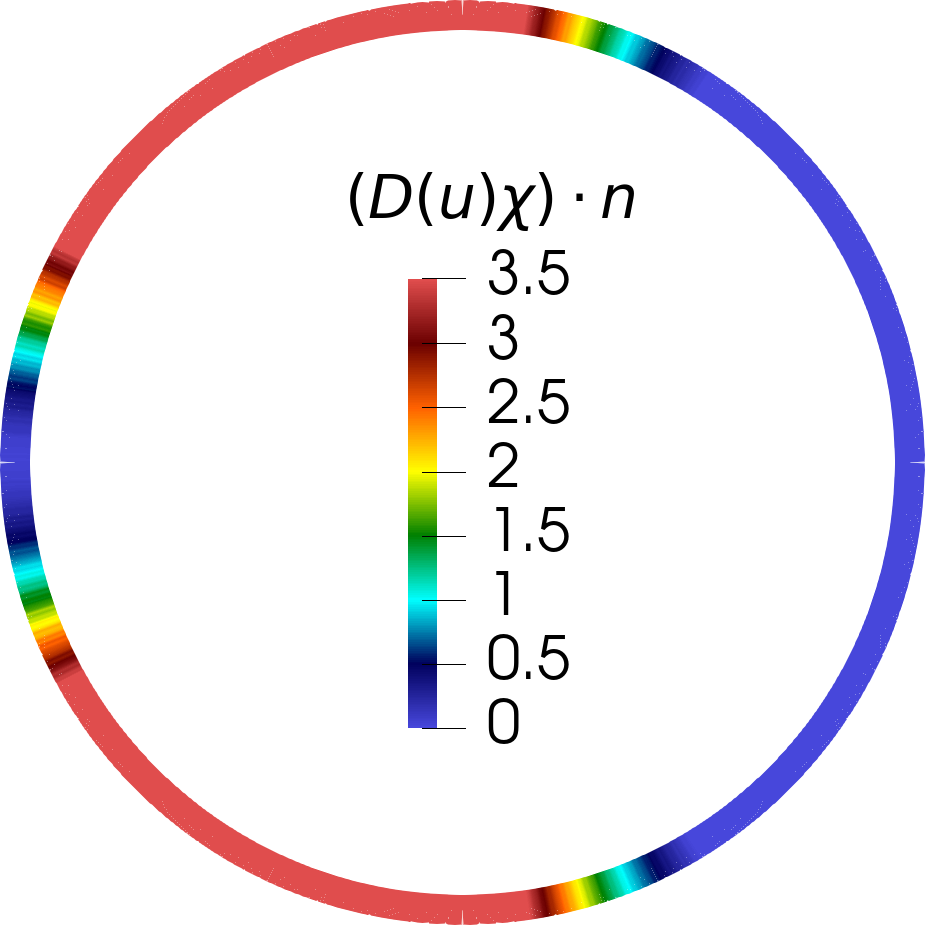}}
\vspace{2mm}
\centerline{\includegraphics[width=1.5in,angle=0]{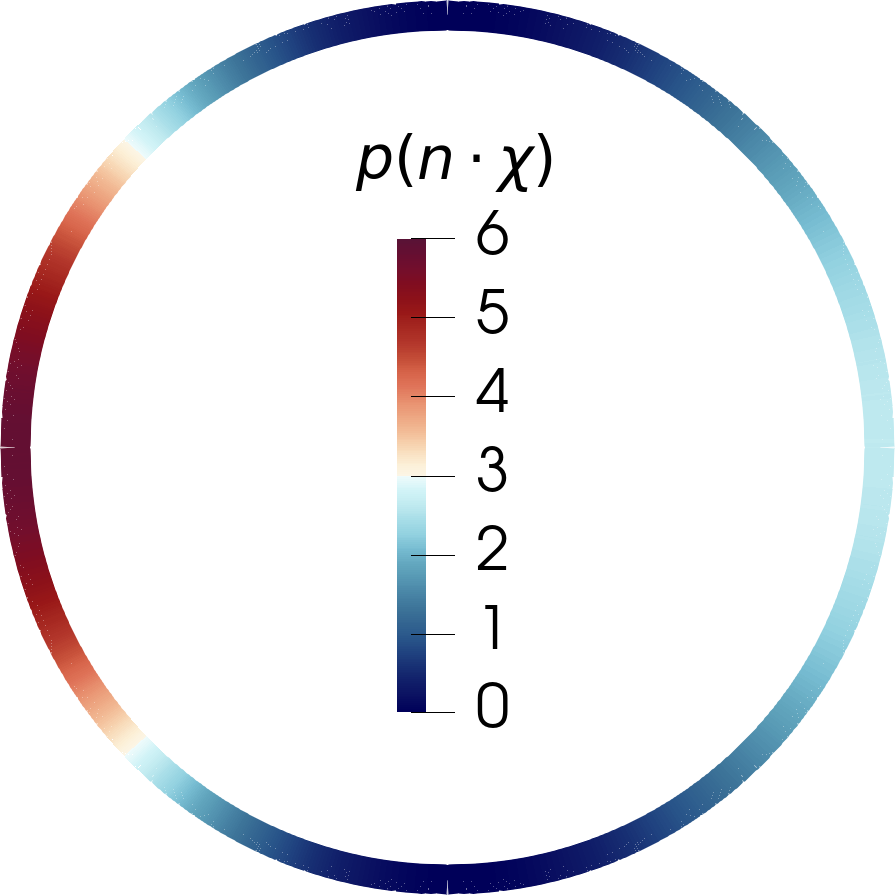}
           \includegraphics[width=1.5in,angle=0]{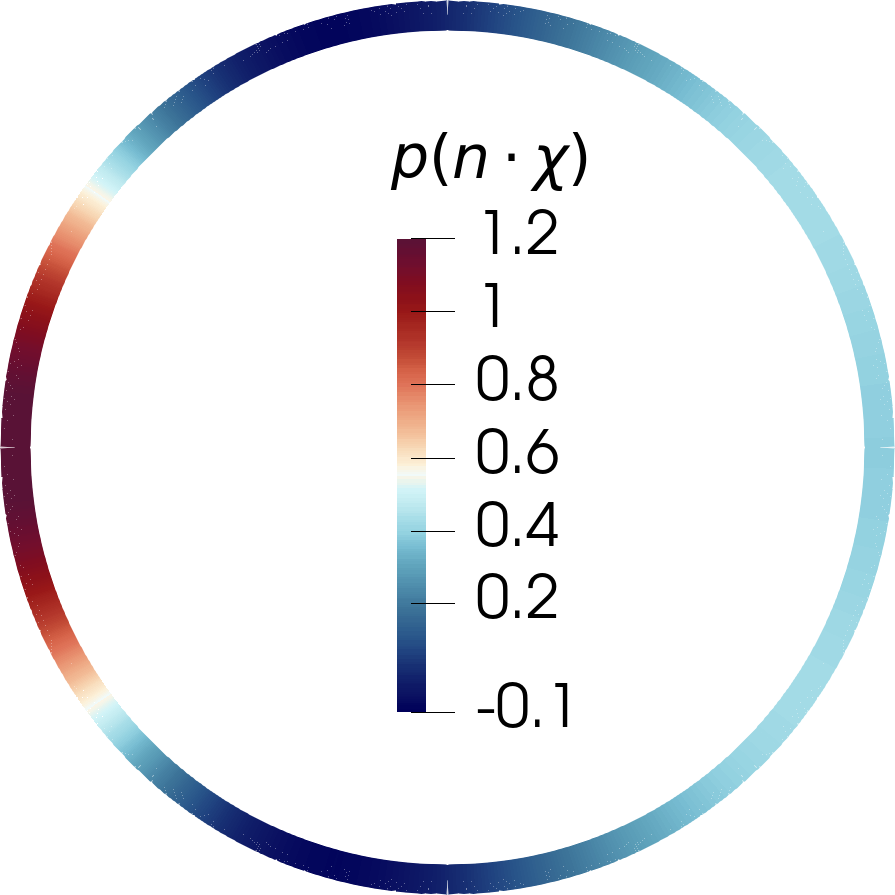}
           \includegraphics[width=1.5in,angle=0]{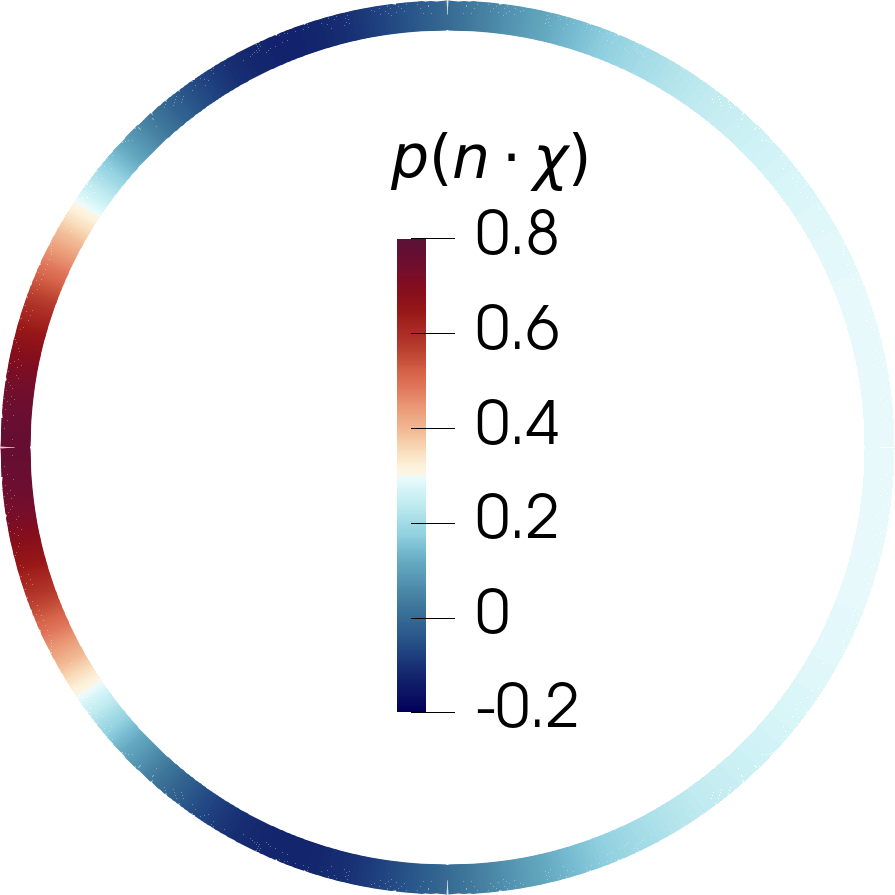}
           \includegraphics[width=1.5in,angle=0]{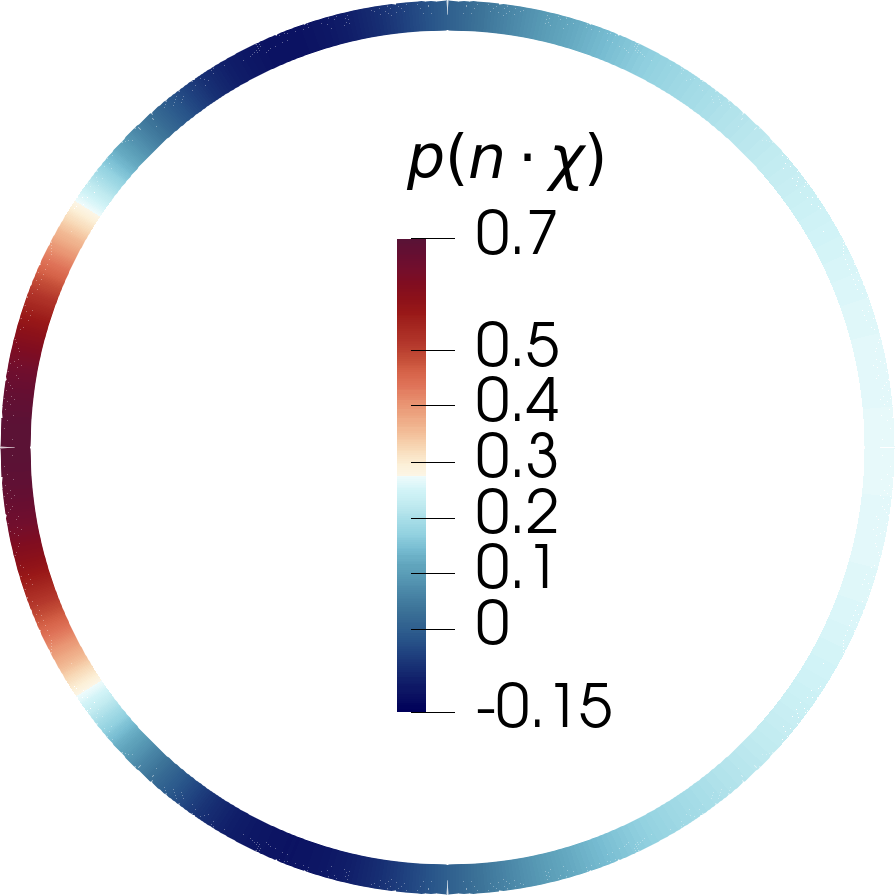}}
\vspace{-2mm}
\caption{Cylinder strain (top) and pressure (bottom)
at $R=1,10,50,100$ (left to right).
Computed using P2-P0.}
\label{fig:ptwopzero}
\end{figure}

Table \ref{tabl:ptoopz} examines the P2-P0 method, presenting percentage
errors using the  Taylor--Hood values on a fine mesh as reference.
The key feature of P2-P0 is that it achieves local mass conservation
(Scott--Vogelius provides mass conservation everywhere).
But the limited accuracy of the drag for larger Reynolds numbers suggests
that this benefit is not worth the cost.
This behavior is consistent with the error estimate \cite[(1.5)]{neilan2020stokes}.
In fact, for Reynolds numbers 50 and 100, the errors actually increase as
the mesh size is reduced.
Thus we suggest not using P2-P0 for sensitive flow simulations, especially
when drag or wall shear stress is required.
A plot of the cylinder strain and pressure values computed with P2-P0
is given in Figure \ref{fig:ptwopzero}.

Table \ref{tabl:crpeeo}
examines the CR-P0 method, presenting percentage
errors using the  Taylor--Hood values on a fine mesh as reference.
The key feature of CR-P0 is that it achieves local mass conservation,
but the velocity is discontinuous.
The main error is in the viscous stress, opposite from P2-P0.

Note that Table \ref{tabl:ptoopz} shows that verification comparisons using
the approach in section \ref{sec:dragcompk} for P2-P0 do not give an
error estimator for drag.
There is strong internal consistency for P2-P0, but the actual accuracy
of the drag values is poor.

Table \ref{tabl:mismat} shows some results of computation of
functionals \eqref{eqn:linfuncipa} and \eqref{eqn:sublinfunc}
with $\uu_h$ and $p_h$ computed with strongly applied boundary
conditions on $\Gamma_h$ and applied to
the approximate Stokes solution $\bchi_h$ for various mesh size parameters $N_\Omega$
and Reynolds numbers $R$ for a channel half-width $W=6.4$.
In the case of strongly applied boundary conditions on $\Gamma_h$,
improvement in accuracy occurred for much larger numbers of segments.
For a number of circle segments greater than $1000N_\Omega$, the computed data agreed
with the data shown to the number of digits reported.
It is notable that in all cases the optimal number of circle segments $N_\Gamma$
satisfied $N_\Gamma=1000N_\Omega$, in the sense that for smaller $S$ the results were
still changing but for larger $S$ the results were identical for the
number of digits presented here.

Table \ref{tabl:mismat} demonstrates the usefulness of our verification
technique in identifying a flawed simulation method.
Unlike P2-P0, which is a method with poor accuracy, Scott--Vogelius
with strongly imposed boundary conditions yields misleading predictions,
with the errors being of order one.
A comparison of Table \ref{tabl:mismat} and Table \ref{tabl:ptoopz} shows
that the pressure drag is not terrible, but the viscous drag is in error
by a factor of 4.
A plot of the erroneous strain values is given in Figure \ref{fig:strainbug}.
We examine this failure in detail in section \ref{sec:constrane}.

There are many other methods that have been proposed for solving the Narier--Stokes
equations. Here we have chosen only four well-established methods to illustrate our approach
to code verification. We mention just two recently proposed methods for which there may be some special benefit for computing wall shear stress on curved domains.

\begin{remark}[Other finite element methods]
In one set of papers \cite{gopalakrishnan2020mass,gopalakrishnan2020weak}, a method
is studied that directly approximates the strain as a primary variable.
Such an approach may be useful for computing wall shear stress and drag
on curved boundaries.

Another direction of interest involves methods for directly approximating problems
on curved domains \cite{bertrand2016parametric,liu2022divergence,oyarzua2019high}.
Since they have higher intrinsic accuracy, such methods may provide
better accuracy for strains on the curved boundary.
\end{remark}

\begin{remark}[Consistency check for time-dependent flows]
A reference computation has been reported in \cite{ref:LinkeRebholzSVconnection,
ref:refvalcyliftdragVolkerJohn} that has been explored in subsequent work \cite{batugedara2022note,
ref:VacekSvacekSVTHP1P1}. The problem considered consists of a short, narrow domain with pulsatile
inflow and outflow. Unfortunately, this problem does not compare closely to a physical problem
that would allow validation of the results.

In \cite{ref:LinkeRebholzSVconnection}, it is shown that grad-div stabilized
Taylor--Hood computations converge to Scott--Vogelius as the penalty
parameter is increased. Unfortunately, it is assumed in \cite{ref:refvalcyliftdragVolkerJohn}
that $\omega=\beta$ for time dependent flows.
The analogous relation for such flows is \cite[section 2.3]{ref:timedepdragformula}
\begin{equation} \label{eqn:timedep}
\beta(\vv)=\omega(\vv)+\int_\Omega \uu_t\,\vv\,d\xx.
\end{equation}
This formula for drag also appears in \cite[4.4]{ref:statsNavStoManica}.
It is also shown in \cite[section 2.3]{ref:timedepdragformula} how to modify
the integration-by-parts formula for grad-div stabilization.
Provided a good approximation of $\uu_t$ is available, the relation
\eqref{eqn:timedep} could be used for verification in the time-dependent case.

Error estimates for drag computation are presented in \cite{ref:errestdragtimedep},
but not including errors due to the approximation of a curved boundary.
\end{remark}

\subsection{Polygonal boundary approximation challenges}
\label{sec:polybac}

In the previous section we found that Scott-Vogelius elements combined with a Dirichlet boundary condition on the cylinder causes spurious
drag values.
As we will see in this section, this is due to an over-constraint of the solution at the vertices of the mesh, which is further caused by the approximation $\Gamma_h$ of the curved boundary of the cylinder.

The section will proceed as follows.
We start by defining in more detail the finite element approximation of the Navier--Stokes equations with Dirichlet boundary conditions.
In Section \ref{sec:testokes}, we show that using this boundary condition yields $H^1(\Omega_h)$-error rates for $\uu_h$ of order $h^\frac{1}{2}$.
This is a cause of concern as the drag evaluation involves evaluating $\nabla \uu_h$ on the boundary of the cylinder.
Formally, this will reduce the accuracy by order 1/2.
In Section \ref{sec:constrane}, we show a counter-example for drag convergence for exactly-divergence free elements combined with Dirichlet boundary conditions on the cylinder.
The counter-example sheds light on the failed drag computations previously reported in Table \ref{tabl:badmat}.

Let us assume that $\partial\Omega\backslash\Gamma$ is polygonal
and that $\Gamma$ is the curved part of the boundary
where we make the polygonal approximation.
Thus the boundary condition $\uu=\gbc$ on $\partial\Omega\backslash\Gamma$
can be implemented using an interpolant $\gbc_I$.

Denote by $\Gamma_h$ this approximation and by $\Omega_h$ the
corresponding domain approximation of $\Omega$.
Define
\begin{equation} \label{eqn:approxom}
a_h(\uu,\vv)=\frac12\int_{\Omega_h} \du:\dv\,d\xx,\quad \dv=\nabla\vv+\nabla\vv^t,
\quad b_h(\vv,p)=\int_{\Omega_h}p\,\sdiv\vv\,d\xx.
\end{equation}
Let $V_h$ denote piecewise polynomials of degree $k$ that vanish on $\Gamma_h$
and on $\partial\Omega\backslash\Gamma$.
Let $\Pi_h$ denote a corresponding pressure space of piecewise polynomials
of degree $k-1$ on $\Omega_h$.
Let $\uu_h^0\in V_h+\gbc_I$ and $p_h\in\Pi_h$ denote the solution of
\begin{equation} \label{eqn:stokvarm}
a_h(\uu_h^0,\vv)+b(\vv,p_h)=0\quad\forall\vv\in V_h,\qquad
b(\uu_h^0,q)=0\quad\forall q \in \Pi_h.
\end{equation}

For Taylor-Hood, we take $k=2$ and $\Pi_h$ to be continuous linear functions.
The P2-P0 method also has $k=2$ for $V_h$ but $\Pi_h$ consists of piecewise constants.
For the Scott--Vogelius method \cite{lrsBIBih}, we take $k=4$
and $\Pi_h=\sdiv V_h$.
In this case, the equations \eqref{eqn:stokvarm} can be solved by
the iterated penalty method \cite{lrsBIBih}.
Computational errors for this method are given in Table \ref{tabl:zerobc}
for the test problem in Section \ref{sec:testokes}.

\begin{table}
\begin{center}
\begin{tabular}{|c|c|c|c|c|}\hline
meshsize $N_\Omega$ & 8 & 16 & 32 & 64 \\
\hline
$H^1(\Omega_h)$ error& 2.23e-01 & 1.58e-01 & 1.11e-01 & 7.19e-02  \\
\hline
error rate $r$ ($h^r$) &  & 0.497 & 0.509 & 0.637 \\
\hline
\end{tabular}
\end{center}
\vspace{-5mm}
\caption{Errors as a function of meshsize for the test problem in
Section \ref{sec:testokes} using the Scott--Vogelius method
with $k=4$ on a mesh generated by {\tt mshr} with a meshsize parameter $N_\Omega$
using a number of circle segments $N_\Gamma=32 N_\Omega$,
imposing the Dirichlet condition $\uu_h^0=\bfz$
on $\Gamma_h$ strongly.
}
\label{tabl:zerobc}
\end{table}

\subsection{Error rates using Dirichlet vs. Nitsche boundary condition}
\label{sec:testokes}

Consider the Stokes equations
\begin{equation}\label{eqn:jussto}
-\Delta\uu+\nabla p=\ff\;\hbox{in}\;\Omega,\qquad
\sdiv\uu=0\;\hbox{in}\;\Omega,\qquad \uu=\gbc\;\hbox{on}\;\partial\Omega,
\end{equation}
on the domain
\begin{equation}\label{eqn:boxcirl}
\Omega=\set{(x,y)}{-\ell_1 <x<\ell_2,\quad -W<y<W,\quad x^2+y^2>1},
\end{equation}
with $\ell_1=6$, $\ell_2=12$, and $W=6$.
Computational results are given in Table \ref{tabl:zerobc} for a test problem where
\begin{equation}\label{eqn:shearstremfn}
\uu=\curl\psi, \qquad \psi(x,y)=\log\big(\sqrt{x^2+y^2}\big)-\half \big(x^2+y^2\big).
\end{equation}
Being a curl, $\uu$ is divergence free.
It is easy to check that $\Delta\psi=-1$ on $\Omega$ since the $\log$
function is harmonic there.
Thus
$$
\Delta\uu=\curl\Delta\psi=0
$$
in $\Omega$.
Therefore $\uu$ is a solution of the Stokes equations with a constant pressure.
It is a shear flow around the cylinder:
\begin{equation}\label{eqn:shearflotest}
\uu(x,y)=\Big(1-\frac{1}{x^2+y^2}\Big)(-y,x).
\end{equation}
Errors using strong application of the zero boundary condition on
$\Gamma_h$ are given in Table \ref{tabl:zerobc}.

\begin{table}
\begin{center}
\begin{tabular}{|c|c|c|c|c|}\hline
meshsize $N_\Omega$ & 4 & 8 & 16 & 32  \\ 
\hline
$H^1(\Omega_h)$ error& 3.60e-02 &  8.64e-03 & 3.85e-03 & 1.08e-03 \\
\hline
error rate $r$ ($h^r$) &  &2.059 & 1.116 & 1.834 \\ 
\hline
\end{tabular}
\end{center}
\vspace{-5mm}
\caption{Errors as a function of meshsize for the Scott--Vogelius method
with $k=4$ on a mesh generated by {\tt mshr} with a meshsize parameter $N_\Omega$
using a number of circle segments $N_\Gamma=32 N_\Omega$.
This is done by imposing the Dirichlet condition $\uu_h^1=\bfz$
on $\Gamma_h$ weakly via Nitsche's method, with
the parameters in \eqref{eqn:asmnits}
defined as follows: $h$ is the maximum mesh size and $\newgamma=100$.
}
\label{tabl:zeronits}
\end{table}


The low accuracy for the strong imposition of Dirichlet boundary conditions
on $\Gamma_h$, as indicated in Table \ref{tabl:zerobc} for the test problem in
Section \ref{sec:testokes}, appears to of order $h^{1/2}$,
much lower than the order $h^{3/2}$ obtained in the scalar case
\cite{lrsBIBaa,scott1975interpolated}.
The difficulty appears to arise due to the incompressibility constraint
which links the two vector components of the solution.

Having an error $\norm{\uu-\uu_h}_{H^1(\Omega_h)}=ch^{1/2}$, as we found
in Table \ref{tabl:zerobc}, is consistent with the pointwise maximum error
$|\nabla(\uu-\uu_h)|$ being only order one ($h^0$)
near the boundary, that is, no order of accuracy for the derivatives near the boundary.

\subsection{A convergence counterexample: Over-constrained strain}
\label{sec:constrane}

In the previous section, we remarked that the low error rate for $\nabla \uu_h$, obtained when combining an exactly-divergence free approximation with a Dirichlet boundary condition, is an issue of concern with respect to drag computation. In this section, we show that this concern is merited: We give an example showing that, given a certain mesh configuration, the $\nabla \uu_h$ converges to the wrong value. This reveals an erroneous over-constraint of $\uu_h$, which is caused by the approximation $\Gamma_h$ of the cylinder boundary.

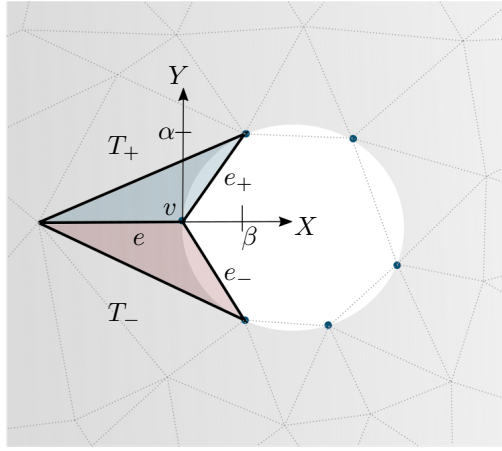
\begin{figure}
\begin{overpic}[scale=1.4]{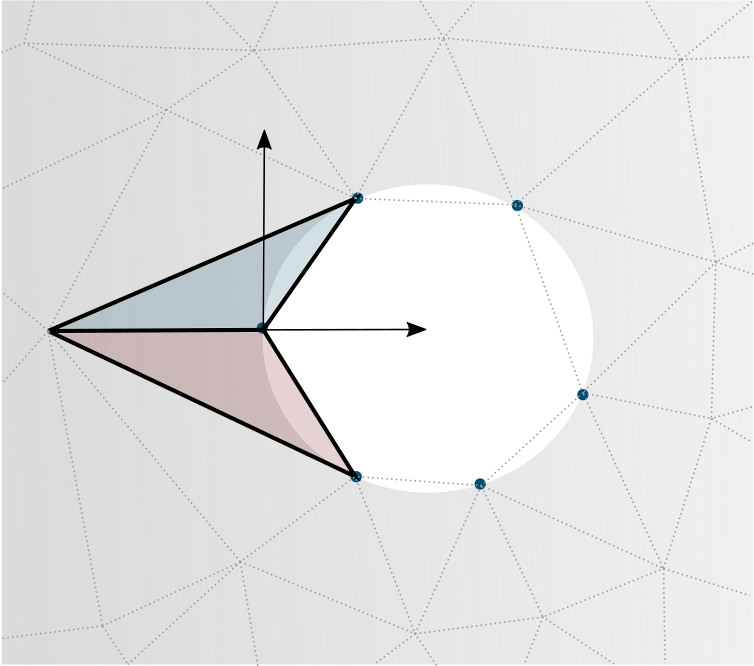}
\put(20,58){\footnotesize $T_+$}
\put(20,25){\footnotesize $T_-$}
\put(31,46){\footnotesize $v$}
\put(25,40){\footnotesize $e$}
\put(43,52){\footnotesize $e_+$}
\put(43,33){\footnotesize $e_-$}
\put(57,42){\footnotesize $X$}
\put(32,72){\footnotesize $Y$}
\put(46,44){\footnotesize $\vert$}
\put(46.5,40){\footnotesize $\beta$}
\put(33,61){\footnotesize $-$}
\put(30,61){\footnotesize $\alpha$}
\end{overpic}
\centering
\caption{Illustration of a geometrical setup which yields spurious drag at vertex $v \in \Gamma_h$.}
\label{fig:illuzor}
\end{figure}

\begin{counterex}

{\rm Geometrical setup.} Consider the case when two triangles $T_{\pm}$ meet at an edge $e$ having  a vertex $v\in\Gamma_h$ such that $e$ is perpendicular to the tangent at $v$. We orient a corresponding coordinate system $(X,Y)$ so that $X$ points in the direction of the normal and $Y$ points in the direction of the tangent.
We assume the two triangles $T_{\pm}$ are symmetric, in the sense that $e_+=(\beta,\alpha)$ and $e_-=(\beta,-\alpha)$.

\noindent \\
{\rm A constraint following from the divergence free condition.} Since $\sdiv\uu_h=0$, we can write $\uu_h=\curl\psi_h$ where $\psi_h$ is a
$C^1$ piecewise polynomial of degree one higher than $\uu_h$ \cite{lrsBIBiu}.
Write $\psi_h^{\pm}$ for the restriction of $\psi_h$ to $T_{\pm}$.
Define $[\;\cdot\;]_e$ to be the jump across $e$.
Since $\psi_h\in C^1$,
\begin{equation*}
\bfz=[\nabla\psi_h]_e=\nabla\psi_h^+|_e-\nabla\psi_h^-|_e,
\end{equation*}
so differentiating the difference in the direction of $e$, we get
\begin{equation*}
\bfz=[\partial_X \nabla\psi_h]_e=\partial_X \nabla\psi_h^+|_e-\partial_X \nabla\psi_h^-|_e.
\end{equation*}
In particular,
\begin{equation}\label{eqn:edifprets}
0=[\partial_X\partial_Y\psi_h]_e=
\partial_X\partial_Y\psi_h^+|_e-\partial_X\partial_Y\psi_h^-|_e.
\end{equation}

\noindent \\
{\rm A constraint following from the Dirichlet boundary condition.}
Let $e_\pm$ be the boundary edges of $T_\pm$ on $\Gamma_h$.
The Dirichlet boundary conditions on $\Gamma_h$ imply that $\nabla\psi_h^{\pm}=\bfz$
on $e_\pm$, so again differentiating zero, we find
\begin{equation}\label{eqn:dernabluts}
\partial_{e_\pm}\nabla\psi_h^{\pm}=\bfz \quad\hbox{on}\;e_\pm.
\end{equation}
In particular, we have
\begin{equation}\label{eqn:moreprets}
\partial_{e_\pm}\partial_X\psi_h^{\pm}(v)=0 .
\end{equation}

For simplicity, assume that we can write
\begin{equation}\label{eqn:assumsits}
\partial_{e_\pm}=\alpha \partial_Y \pm \beta \partial_X.
\end{equation}

We have $\beta\neq 0$ and, if the mesh is fine enough, $\alpha\approx 1$.
Combining \eqref{eqn:moreprets} and \eqref{eqn:assumsits}, we find
\begin{equation}\label{eqn:combosits}
\alpha \partial_Y\partial_X \psi_h^+(v)+ \beta \partial_X^2\psi_h^+(v)=0=
\alpha \partial_Y\partial_X \psi_h^-(v)- \beta \partial_X^2\psi_h^-(v).
\end{equation}
But $\psi_h^\pm=\psi_h$ on $e$, so we can write \eqref{eqn:combosits} as
\begin{equation}\label{eqn:rewertits}
2 \beta \partial_X^2\psi_h(v)=
\alpha \big(\partial_Y\partial_X \psi_h^-(v)- \partial_Y\partial_X \psi_h^+(v) \big).
\end{equation}

\noindent \\
{\rm Combining the two constraints.}
Note that $\partial_Y\partial_X \psi_h^\pm=\partial_X\partial_Y \psi_h^\pm$
because $\psi_h^\pm$ are polynomials.
Applying \eqref{eqn:edifprets} to \eqref{eqn:rewertits}, we get
\begin{equation}\label{eqn:bigfinish}
2 \beta \partial_X^2\psi_h(v)=0\implies  \partial_X^2\psi_h(v)=0,
\end{equation}
as $\beta\neq 0$.
Since $e$ is perpendicular to the tangent at $v$,
$\partial_r=-\partial_X$.
But the exact streamfunction \eqref{eqn:shearstremfn}
satisfies $\partial_r^2\psi=-2$ on $\Gamma$.
So an order one (i.e. $h^0$)
error occurs in the strain at the vertices.

We can interpret the implications of \eqref{eqn:bigfinish} in terms of
$\nabla\uu$ as follows.
Using \eqref{eqn:dernabluts} and \eqref{eqn:assumsits}, we get
\begin{equation}\label{eqn:twodinish}
0=\alpha\psi_{h,YY}(v)\pm\beta\psi_{h,XY}(v), \qquad
0=\alpha\psi_{h,YX}(v)\pm\beta\psi_{h,XX}(v).
\end{equation}
But \eqref{eqn:bigfinish} and the right-hand equation in \eqref{eqn:twodinish}
imply $0=\psi_{h,YX}(v)=\psi_{h,XY}(v)$.
Using the left-hand equation in \eqref{eqn:twodinish}, we
then find $0=\psi_{h,YY}(v)$.
Therefore the full Hessian of $\psi_h$ vanishes at $v$, and thus
we conclude that
\begin{equation}\label{eqn:bigrinish}
\nabla\uu_h(v)=\bfz.
\end{equation}
This constraint will occur at the vertices of any polygonal domain.
\end{counterex}

The above example assumes a certain symmetry of the mesh.
For non-symmetric meshes, the constraint \eqref{eqn:rewertits} can be written using $\alpha_\pm, \beta_\pm$.
We see that, generally, \eqref{eqn:rewertits} expresses a constraint that depends only on the mesh configuration.
Combining it with divergence free elements causes the approximation $\nabla u_h$ to be determined by the mesh configuration (rather than the model equations).
Thus we expect this issue to occur for more general meshes, as was illustrated in Table \ref{tabl:badmat}.
We give a proof of this in Appendix \ref{sec:ttappndx}.

\begin{remark}[Previously reported  constraints related to incompressibility]

The over-constraint of exactly divergence free functions has been
reported before for isogeometric elements \cite[(17)]{buffa2011isogeometric},
in that case causing issues at the corners of a rectangular domain.
This is an example of a singular boundary vertex \cite[(4.2)]{brezzi1991stability}
and \cite[Figure 3]{lrsBIBbj}.

A similar constraint plays a role in defining a novel element \cite{ainsworth2021mass}.
The authors study a Hermite variant of the standard Scott--Vogelius element,
in which the velocities $\vv_h$ are $C^1$ at vertices and the corresponding pressures
$q_h$ are $C^0$ at vertices, with the exception of the vertices of the
polygon boundary.
At such points on the boundary, they relax the continuity constraints by one order
($q_h\in\Pi_h$ are allowed to be discontinuous at such points).
This is needed because having $\vv_h\in V_h$ vanish on the polygonal boundary forces
$\nabla\vv_h$ to be zero there if $\vv_h\in C^1$, and thus $\sdiv\vv_h=0$ there.
But this means that $\sdiv\vv_h$ cannot match arbitrary $q_h\in \Pi_h$, which would be
continuous but not zero, causing a mismatch between the pressure space and $\sdiv V_h$.
\end{remark}

\begin{figure}
\begin{tikzpicture}
   \node[anchor=south west,inner sep=0] at (0,0) {\includegraphics[width=0.9\textwidth]{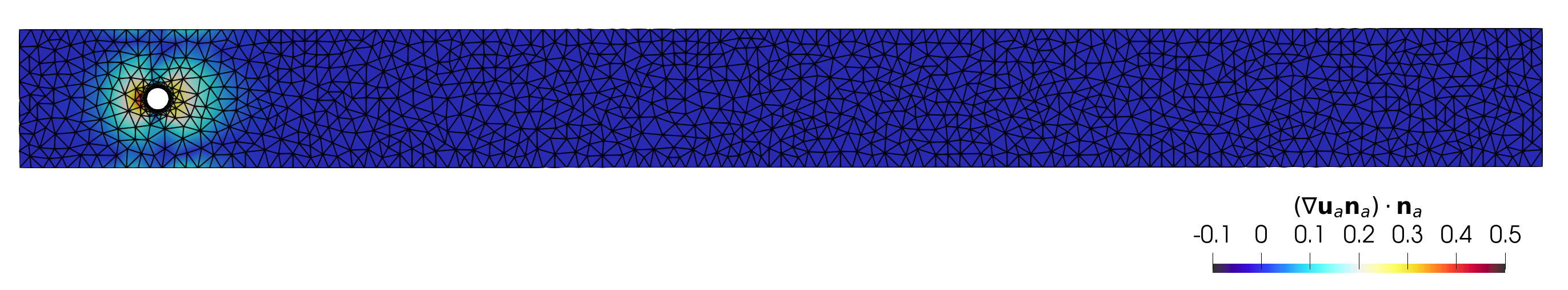}};
   \draw[red,ultra thick] (1.2,1.8) rectangle (1.8,2.1);
	\draw[-stealth,red,ultra thick] (1.2,1.8) -- (0,-0.25);
	\draw[-stealth,red,ultra thick] (1.8,1.8) -- (7,-0.25);
	\draw[black,ultra thick] (8,-0.25) -- (8,-5);
	\draw[black,ultra thick] (8,-0.25) -- (16,-0.25);
	\node[anchor=south west,inner sep=0] at (0,-4) {\includegraphics[width=0.45\textwidth]{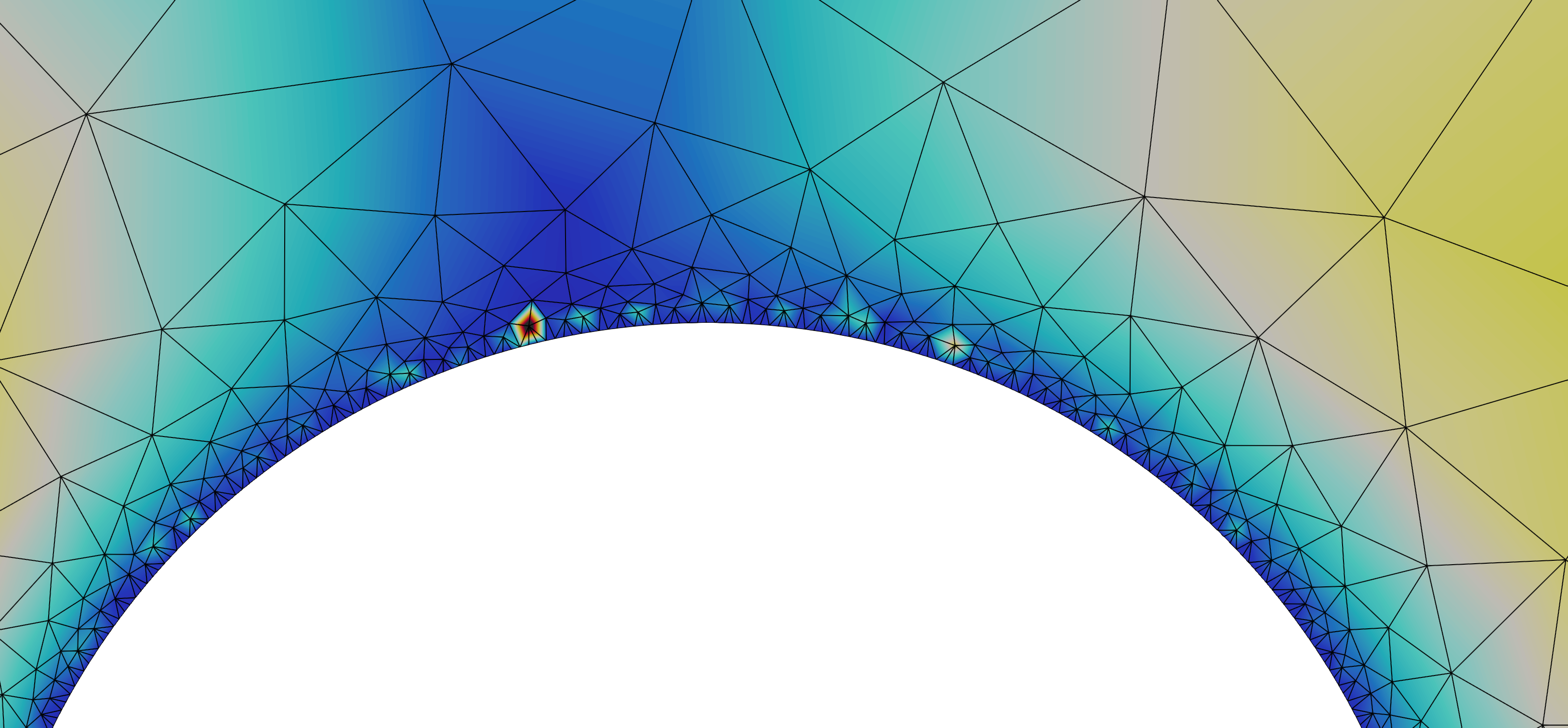}};
		\node[anchor=south west,inner sep=0] at (9,-4) {\includegraphics[width=0.45\textwidth]{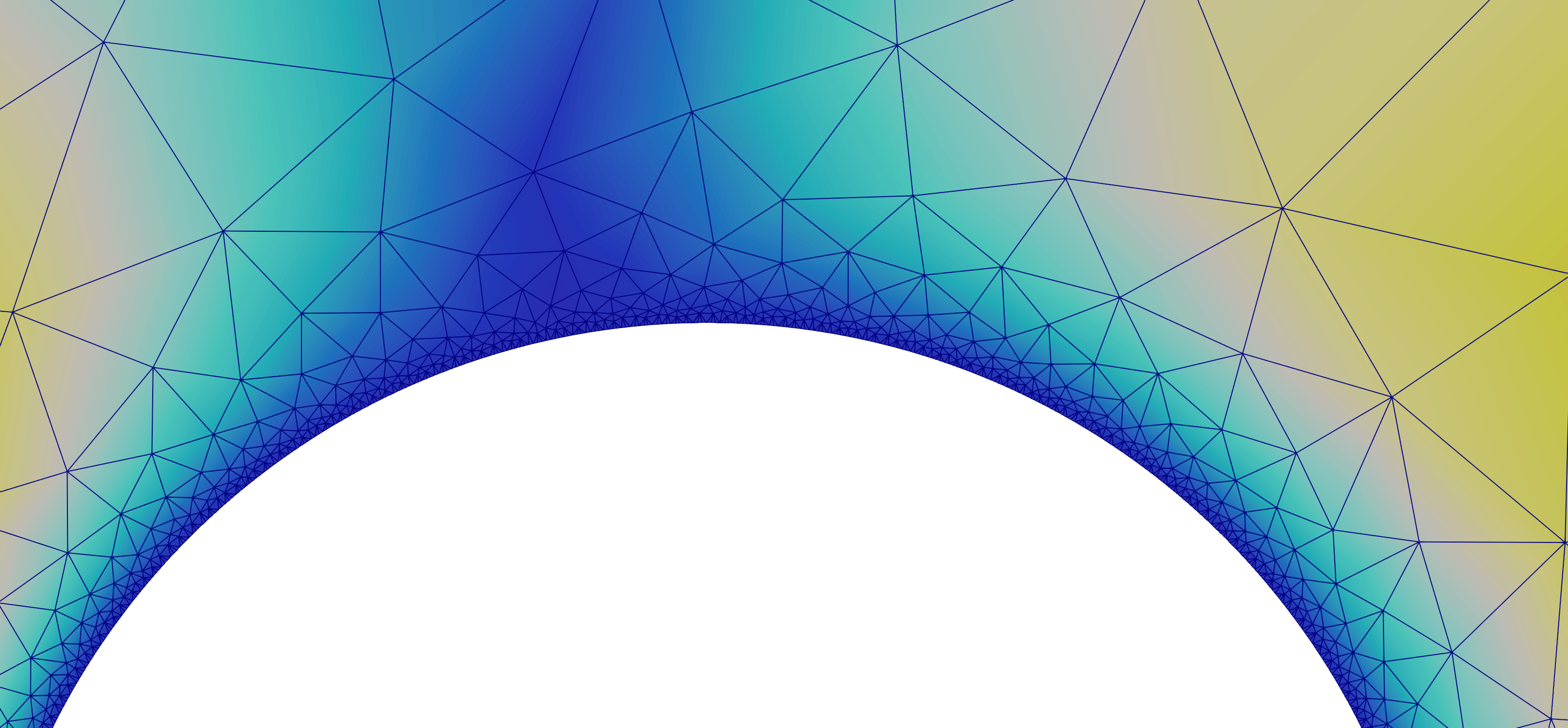}};
\node at (3,-4.5) {\footnotesize Scott Vogelius with no-slip boundary condition};
\node at (12,-4.5) {\footnotesize Taylor Hood with no-slip boundary condition};
\end{tikzpicture}
\caption{Radial strain around the cylinder $(\nabla u_h \nn_a) \cdot \nn_a$ for Scott-Vogelius (bottom left) and Taylor-Hood (bottom right) discretizations, both with no-slip boundary conditions on the cylinder boundary $\Gamma_h$. The former exhibits spurious fluctuations at the cylinder boundary, caused by the over-constraint of the solution on this boundary.}
\label{fig:strainbug}
\end{figure}
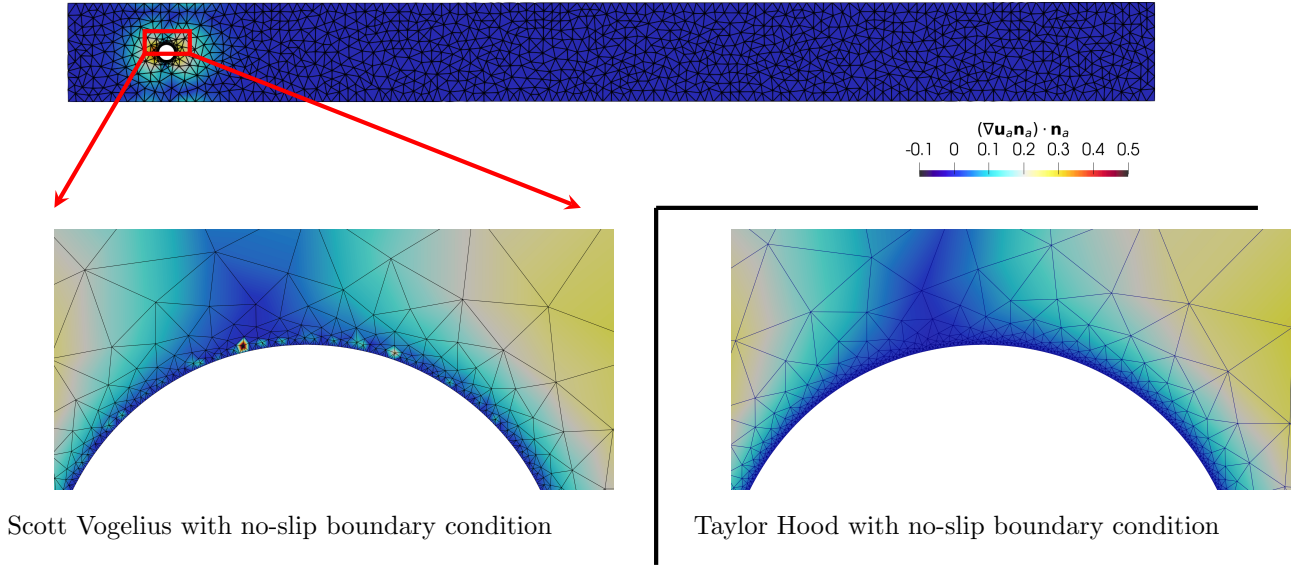

We now consider relaxing the constraint related to the strong
imposition of Dirichlet boundary conditions.

\subsection{Nitsche's method for imposition of no-slip}
\label{sec:testshes}

Nitsche's method is based on the variational form
\begin{equation}\label{eqn:asmnits}
N_h(\uu,\vv)=
a_h(\uu,\vv)-\int_{\Gamma_h}\derdir{\uu}{n}\cdot\vv
+\uu\cdot\derdir{\vv}{n} - \frac{\newgamma}{h}\uu\cdot\vv\,ds ,
\end{equation}
where $\newgamma$ is a parameter that must be chosen sufficiently large.
This imposes the Dirichlet boundary condition on $\Gamma_h$ weakly.

Let $\uu_h^1\in V_h+\gbc_I$ and $p_h\in\Pi_h$ denote the solution of
\begin{equation} \label{eqn:nitstokvarm}
N_h(\uu_h^1,\vv)+b(\vv,p_h)=0\quad\forall\vv\in V_h,\qquad
b(\uu_h^1,q)=0\quad\forall q \in \Pi_h.
\end{equation}

Errors for this method are given in Table \ref{tabl:zeronits}.
They are orders of magnitude smaller than using the strong imposition
of the homogeneous Dirichlet condition on $\Gamma_h$.
The typical choice for the parameter $\newgamma$ was $\newgamma=10^6$.

\subsection{Exact boundary conditions}
\label{sec:polybcs}

One way to assess the potential for improving the accuracy
on polygonal approximations is to use the exact boundary conditions,
which are available for a manufactured test problem as defined
in Section \ref{sec:testokes}.
The errors in this case are presented in Table \ref{tabl:polybcs}.
Thus we see that the Nitsche method already achieves almost all of the
available accuracy for this problem.
Thus the motivation for considering more sophisticated algorithms,
such as \cite{bramble1972projection}, is diminished.

\begin{table}
\begin{center}
\begin{tabular}{|c|c|c|c|c|}\hline
meshsize $N_\Omega$ & 4 & 8 & 16 & 32  \\ 
\hline
$H^1(\Omega_h)$ error& 1.15e-02 &  6.48e-03 & 2.77e-03 & 7.86e-04 \\
\hline
error rate $r$ ($h^r$) &  &0.828 & 1.226 & 1.817 \\
\hline
\end{tabular}
\end{center}
\vspace{-5mm}
\caption{Errors as a function of meshsize for the Scott--Vogelius method
with $k=4$ on a mesh generated by {\tt mshr} with a meshsize parameter $N_\Omega$
using a number of circle segments $N_\Gamma=32 N_\Omega$.
This is done by using the exact Dirichlet condition \eqref{eqn:shearflotest}
extended to $\Gamma_h$.
}
\label{tabl:polybcs}
\end{table}

\section{Validation: Lamb's model}
\label{sec:valamb}

In the last two sections, we showed how verification methods (are you obtaining the right solution?) revealed distinct differences in the drag computation offered by different finite elements.
In this section, we turn our attention to validation (are you getting the right physics?).
Classically, this has been done by comparing experimental results with either simulation results \cite{ref:SchaeferTurek1996} or analytical solutions \cite{ref:cyldraglowRe}.
We here augment this by comparing simulations, experiments and an analytical solution in the form of Lamb's model.

A formula for the drag of an infinite cylinder in an unbounded domain
was derived by Lamb \cite[343., page 614]{lamb1993hydrodynamics} and
takes the form
\begin{equation} \label{eqn:lambodel}
D_L(R)=\frac{g(R)}{R},\qquad g(R)=\frac{8\pi}{\half-\gamma-\log(R/8)},
\end{equation}
where $\gamma\approx 0.57721566490$ is Euler's constant and $R$ is the Reynolds number.
This was proposed as a model only for small $R$, and it has a singularity
when $R=8e^{-\gamma+\half}\approx 7.4$.

Previously, experimentally measured drag \cite{ref:cyldraglowRe} has been shown to deviate substantially from the drag predicted from Lamb's model. The deviation is monotonically increasing as $R$ decreases, as indicated in Table \ref{tabl:datdisc}. By examining this computationally, we find the discrepancy can be explained by effects due to domain width.
In Table \ref{tabl:vardub}, we examine the dependence of the drag coefficient on the domain width $W$. For small Reynolds numbers, the drag coefficient changes significantly as a function of $W$, and it is clear that a very large value of $W$ is needed to obtain a drag value close to the prediction of Lamb.
For larger values of Reynolds numbers, where the Lamb model is not applicable,
the drag values vary less as a function of $W$.

\begin{table}
\begin{center}
\begin{tabular}{|c|c|c|c|c|}\hline
$R$ & $D_L$ & $D_e$ & $D^a_s$ & $D^b_s$ \\
\hline
2.636  &  9.23 &  9.14 & 6.80 &  6.09 \\
0.9741 & 12.72 & 13.99 & 13.93& 11.89 \\
0.6479 & 15.78 & 22.93 & 19.45& 16.12 \\
0.2274 & 31.73 & 53.94 & 50.40& 38.50 \\
\hline
\end{tabular}
\end{center}
\vspace{-5mm}
\caption{Drag data for low Reynolds numbers.
The quantity $D_L(R)$ is Lamb's model \cite[(3)]{ref:fallingcylinderdrag}
or \cite[(1)]{ref:Finn53lowRecyldrag}.
The experimental data $D_e(R)$ is the quantity $D$
from \cite[Table III]{ref:cyldraglowRe}, computed by multiplying the
penultimate column by $D_L$.
The simulation data $D^x_s$ ($x=a$ or $x=b$) was computed as in
section \ref{sec:testshes}
for domains \eqref{eqn:boxcirl} with (a) $\ell_1=12.8$, $\ell_2=128$, $W=64$
and (b) $\ell_1=30$, $\ell_2=300$, $W=150$, using Scott--Vogelius--Nitsche
with $N_\Omega=128$, $N_\Gamma=1024$, and $\newgamma=10^6$.
}
\label{tabl:datdisc}
\end{table}

\subsection{Data discrepancy in \cite{ref:cyldraglowRe}}

\begin{table}
\begin{center}
\begin{tabular}{|r|c|c|c|c|c|c|c|c|c|}\hline
$R\;\backslash\; W$ &
     6.0   &  6.4  &   8    &   10   &    15   &   20   &   30  & 150 & Lamb \\
\hline
0.01&2723.9 & 2560.1 &2111.6 & 1791.6 &  1417.6 & 1268.7 & 1167.3 & 1119.8 & 380.37 \\
0.1 &272.43 & 256.06 &211.23 & 179.26 &  141.93 & 127.12 & 117.12 & 112.64 & 58.383 \\
1 & 27.672 & 26.074 & 21.753 & 18.764 &  15.539 & 14.448 & 13.849 & 13.667 & 12.552 \\
10 & 4.240 &  4.097 &  3.717 &  3.468 &   3.229 &  3.161 &  3.129 &  3.121 & NA \\
50 & 1.918 &  1.866 &  1.726 &  1.634 &   1.548 &  1.524 &  1.515 &  1.513 & NA \\
100& 1.513 &  1.471 &  1.355 &  1.278 &   1.202 &  1.181 &  1.173 &  1.173 & NA \\
\hline
\end{tabular}
\end{center}
\vspace{-5mm}
\caption{Computation of $\omega$ for various channel half-widths $W$ computed
for the domain \eqref{eqn:oneomega}
using Scott--Vogelius--Nitsche as in section \ref{sec:testshes}, with
$N_\Omega=128$, $N_\Gamma=1024$, and $\newgamma=10^6$.
The last column gives the drag from Lamb's model for $R\leq 1$.
For larger $R$, Lamb's model is not applicable, and we have indicated this with NA.
}
\label{tabl:vardub}
\end{table}

Table \ref{tabl:datdisc} shows the discrepancy highlighted in \cite{ref:cyldraglowRe}.
The data discrepancy can possibly be explained by boundary effects.
We see that the drag data in \cite{ref:cyldraglowRe} agrees well with
simulation data done for a fairly wide ($W=64$) domain.
However, it is seen in Figure \ref{fig:potential} that choosing
an even wider ($W=150$) domain yields simulation data in
closer agreement with the Lamb model \eqref{eqn:lambodel}.
Note that the wider domain was also taken to be longer ($\ell_2$), and
with a longer inflow buffer ($\ell_1$), to keep
the same aspect ratio for the computational domain.

The explanation for the discrepancy in \cite{ref:cyldraglowRe} has been
given before \cite{ref:confincyliflowfvolume,ref:hunerrorkhalil}.
However, the explanation in \cite[page 375]{ref:hunerrorkhalil} is used
to criticize the experimental design in both \cite{ref:cyldraglowRe}
and \cite{ref:fallingcylinderdrag} equivalently.
But the data in \cite{ref:fallingcylinderdrag} is featured in
Figure \ref{fig:potential}, and our assessment of that data
suggests it is quite accurate, as shown in Figure \ref{fig:potential}.
Thus we hope that our simulations can add some clarity to the discussion.

Note that we chose a very large domain in Figure \ref{fig:potential}, both to
insure good agreement with Lamb and to demonstrate the power of the software
used to deal with very large aspect ratios.

Our computations support the validity of Lamb's model,
which we can express analytically as
$$
\lim_{r\to\infty}D_r(R) R=g(R)+o(g(R)),\; R>0.
$$
Several enhancements to Lamb's model are described in
\cite{ref:confincyliflowfvolume} that suggest what the error
term $o(g(R))$ might be.
The effect of a finite domain is often called a blockage effect
\cite{ref:confincyliflowfvolume,chen1995bifurcation,ref:blockagecylinderatio,ref:lowReblocylindr}.
See \cite{ref:confincyliflowfvolume} for various models accounting
for finite domain sizes.

\subsection{Lamb's model and the Stokes Paradox}

In the previous section, we saw that Lamb's model holds for small $R$.
It is interesting to ask if there is an independent way to know that a
correction term $g>0$ is appropriate.
Note that
$$
\lim_{R\to 0} g(R)=0.
$$
It is plausible that a relation such as \eqref{eqn:lambodel} does hold
for some function $g$ that tends to zero as $R\to 0$.
For if $g(R)=RD(R)$ tended to a positive limit at $R=0$, that is $g(0)>0$,
this would imply that
we could determine the drag in terms of a Stokes problem for the infinite domain.
But this would violate Stokes' Paradox \cite{lrsBIBjq}, as we now describe.

The reasoning regarding the Stokes Paradox is as follows.
We can define the drag $D_W(R)=\beta(\bchi)$ on a domain $\Omega_W$
of radius $W$ for any $R$ using the drag formula \eqref{eqn:linfuncipa}.
For $R>0$, it is plausible \cite{ref:Finn53lowRecyldrag} that the limit
$\lim_{W\to\infty} D_W(R)$ exists.
Dividing \eqref{eqn:navstokipa} by $\nu=1/R$, we get an equation
\begin{equation} \label{eqn:aprxnavstok}
-\Delta\uu +R\uu\cdot\nabla\uu+\nabla q =\bfz,\qquad \sdiv\uu=0
\end{equation}
where $q=p/\nu$.
The definition of $\beta$ becomes
\begin{equation} \label{eqn:newdefbeta}
\beta(\vv) =\nu\oint_{\bog} \big((\du -qI)\vv\big)\cdot\nn \,ds,
\end{equation}

The quantity $D_r(R)R$ is an integral
\begin{equation} \label{eqn:nuerdefbeta}
D_W(R)R=\beta(\bchi)R =\oint_{\bog} \big((\du -qI)\bchi\big)\cdot\nn \,ds,
\end{equation}
involving the approximate Stokes problem \eqref{eqn:aprxnavstok}
on $\Omega_W$, and thus the limit
$$
\lim_{R\to 0} D_W(R) R
=\oint_{\bog} \big((\du^W -q^WI)\bchi\big)\cdot\nn \,ds
$$
would converge for each $W$, where $\uu^W,q^W$ is the solution of the Stokes problem
\begin{equation} \label{eqn:limxnavstok}
-\Delta\uu^W +\nabla q^W =\bfz,\qquad \sdiv\uu^W=0
\end{equation}
on $\Omega_W$.
But we would not expect
$$
\lim_{W\to\infty}\lim_{R\to 0} D_W(R) R
$$
to be a nonzero constant since $\lim_{W\to\infty}\uu^W$ is not well defined due
to the Stokes Paradox.
Indeed, Lamb used the Oseen resolution of the Stokes paradox
\cite{lrsBIBjq,lamb1993hydrodynamics} to derive the model \eqref{eqn:lambodel}.

\section{Conclusions}

We have demonstrated that computational simulations can be used to
substantiate Lamb's model for cylinder drag at low Reynolds numbers
and to address proposed disagreements with that model.
We used two formulas for drag to provide an internal check on
the accuracy of the computations.
In the process, we discovered some subtle computational issues with
regard to curved boundaries.
However, we showed that using Nitsche's method to impose boundary
conditions on the cylinder restored high accuracy and
allowed accurate computation of drag.
The use of Nitsche's method to impose boundary
conditions on curved boundaries may be of use in many other contexts.

Our approach to verification and validation of drag can, and should, be
used to assess all codes used to simulate wall shear stress.

\section{Acknowledgments}


We thank Tabea Tscherpel for discussions on the artificial constraints for exactly
divergence-free finite element functions on 2D polyhedral domains,
see \eqref{eqn:bigrinish}.
This has led to the proof in the general case, cf. Appendix \ref{sec:ttappndx}.

\appendix

\section{Proof of Lemma \ref{lem:stokinbypa}}
\label{sec:proofstok}

There are two parts to \eqref{eqn:stokinbypa}.
Both are a consequence of the divergence theorem
\begin{equation} \label{eqn:divtheoypa}
\intox{\sdiv\ww}=\oint_{\bo} \ww\cdot\nn\,ds.
\end{equation}
For the pressure term, we apply this to $\ww=p\vv$.
Since $\sdiv(p\vv)=\nabla p\cdot\vv+p\,\sdiv\vv\,$,
we find that
$$
\intox{\nabla p\cdot\vv+p\sdiv\vv}=\oint_{\bo} p \, \vv\cdot\nn\,ds,
$$
which gives the desired identity for the pressure terms.

We now apply the divergence theorem to  $\ww=\du \vv$.
We first prove that
\begin{equation} \label{eqn:claimnbypa}
\sdiv(\du \vv)=(\Delta\uu)\cdot\vv +\vv\cdot\nabla(\sdiv\uu)+ \half \du:\dv
\end{equation}
by expanding using indices:
\begin{equation} \label{eqn:edrvenbypa}
\begin{split}
\sdiv(\du \vv)&=\sum_i (\du \vv)_{i,i}
=\sum_{ij} (\du_{ij} \vv_j)_{,i}
=\sum_{ij} \big(\du_{ij} \vv_{j,i} + \du_{ij,i} \vv_j\big) \\
&=\sum_{ij} \du_{ij} \vv_{j,i}
+ \sum_{ij}  (\uu_{i,ji} +\uu_{j,ii})\vv_j \\
&= \du:\nabla\vv^t +  \vv\cdot\nabla(\sdiv\uu)  +\Delta\uu\cdot\vv .
\end{split}
\end{equation}
But the symmetry of $\du$ implies that
$$
\du:\nabla\vv^t = \du^t:\nabla\vv = \du:\nabla\vv
$$
so that $\du:\dv=\du:\nabla\vv^t + \du:\nabla\vv = 2\du:\nabla\vv^t$.
Thus
$$
\sdiv(\du \vv)=\half \du:\dv +  \vv\cdot\nabla(\sdiv\uu) +  \Delta\uu\cdot\vv,
$$
proving \eqref{eqn:claimnbypa}.
A second application of the divergence theorem confirms \eqref{eqn:stokinbypa}.

\section{Proof of \eqref{eqn:bigrinish} in the general case}
\label{sec:ttappndx}

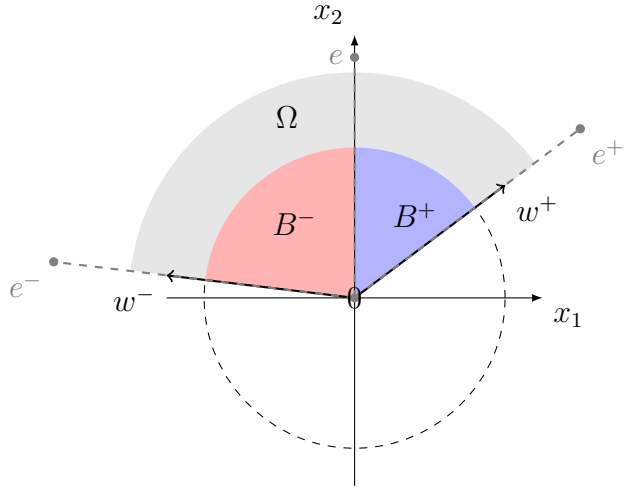
\begin{figure}
\begin{center}
\begin{tikzpicture}
			\draw[dashed] (0, 0) circle [radius=2];
			
			\fill[gray!20] (0, 0) -- (36.6:3) arc (36.6:173:3) -- cycle;
			\fill[blue!30] (0, 0) -- (36.6:2) arc (36.6:90:2) -- cycle;
			\fill[red!30] (0, 0) -- (90:2) arc (90:173:2) -- cycle;
			
			\node at (-0.9, 2.4) {$\Omega$};
			
			\draw[-latex] (-2.5, 0) -- (2.5, 0) node[below right] {$x_1$};
			\draw[-latex] (0, -2.5) -- (0, 3.5) node[above left] {$x_2$};
			
			\coordinate (v1) at (2, 1.5);
			\coordinate (v2) at (-2.5,0.3);
			
			\draw[->,thick] (0, 0) -- (v1) node[below right] {$w^+$};
			\draw[->,thick] (0, 0) -- (v2) node[ below left] {$w^-$};
			
			\draw[gray,dashed,thick] (0, 0) -- (3,2.25) node[below right] {$e^+$};
			\draw[gray,dashed,thick] (0, 0) -- (-4,0.48) node[below left] {$e^-$};
									\draw[gray,dashed,thick] (0, 0) -- (0,3.2) node[ left] {$e$};
						
			\draw[gray,fill=gray] (3,2.25) circle (.3ex);
			\draw[gray,fill=gray] (0,0) circle (.3ex);
			\draw[gray,fill=gray] (-4,0.48) circle (.3ex);
	    	\draw[gray,fill=gray] (0,3.2) circle (.3ex);
				
			\node at (0.8, 1.1) {$B^+$};
			\node at (-0.8, 1) {$B^-$};
			
			\node at (0, 0) {$0$};
\end{tikzpicture}
\end{center}
\caption{Edges meeting at a vertex of the polygon boundary.}
\label{fig:domain-2D}
\end{figure}

Consider a bounded polyhedral domain $\Omega \subset \mathbb{R}^2$.
Without loss of generality we assume that $\bfz$ is contained in the boundary
of $\Omega$, that the domain has an interior angle $\neq \pi$,
and that the point $(0,1)^t$ is contained in the interior of $\Omega$,
cf.~Figure \ref{fig:domain-2D}. 

This means that there are linearly independent vectors $w^\pm$ representing the
directions of the edges $e^\pm$, where
\begin{align*}
	w^+ = \begin{pmatrix} 1 \\ \alpha^+ \end{pmatrix} \quad \text{ and }
	\quad 	w^- = \begin{pmatrix} -1 \\ \alpha^- \end{pmatrix},
\end{align*}
with $\alpha^\pm \in \mathbb{R}$
Linear independence of $w^+$ and $w^-$ is equivalent to the fact that $\alpha^+ + \alpha^- \neq 0$.
Furthermore, we assume that the triangulation of $\Omega$ contains an edge $e$ that contains $0$, and an additional vertex on the positive $x_2$-axis. \
The two parts of the neighborhood we denote the open sets 
\begin{align*}
		B^- \coloneqq B_{\epsilon}(0) \cap \{\xx \in \Omega \colon x_1 < 0\},\\
	B^+ \coloneqq B_{\epsilon}(0) \cap \{\xx \in \Omega \colon x_1 > 0\}.
\end{align*}
The boundary edges associated with the vectors $w^\pm$ we denote by $e^\pm$.

We assume that $\uu \colon \overline{\Omega} \to \mathbb{R}^2$, with
\begin{align*}
\xx = \begin{pmatrix}
		x_1\\x_2
	\end{pmatrix}
\mapsto
	\uu(\xx) = \begin{pmatrix}
	u_1 (\xx)\\u_2(\xx)
\end{pmatrix}
\end{align*}
globally Lipschitz continuous on $\overline{\Omega}$ and piecewise $C^1$ on the triangulation $\mathcal{T}$ of $\Omega$.
In particular there are Lipschitz continuous functions $u^\pm$ defined on $\overline{B}^\pm$ respectively such that
\begin{align*}
	\uu(\xx) = \begin{cases}
		\uu^+(\xx) \quad & \text{ for } \xx \in B^+,\\
				\uu^-(\xx) \quad &  \text{ for } \xx \in B^-.
	\end{cases}
\end{align*}
Since $\uu$ is globally Lipschitz continuous, its gradient $\nabla \uu$ is an
$L^\infty$ function, piecewise defined as
\begin{equation*}
\nabla \uu^\pm(\xx) = \begin{pmatrix}
	\partial_{x_1} u_1^\pm(\xx) & 	\partial_{x_2} u_1^\pm(\xx) \\
		\partial_{x_1} u_2^\pm(\xx) & 	\partial_{x_2} u_2^\pm(\xx)
\end{pmatrix}	
\end{equation*}
on $B^\pm$, respectively.
Since $\uu$ is piecewise $C^1$, the gradients $\nabla \uu^\pm$ are piecewise continuous on neighborhoods of the edges $e^\pm$ and $e$ in $\Omega$.

Let us use the following facts on $\uu$:
\begin{enumerate}
	\item \label{itm:div} $\uu$ is exactly divergence-free, i.e. $\mathrm{tr}(\nabla u^\pm) = 0$, on $T^\pm$, or
	\begin{align*}
		\partial_{x_1} u_1^\pm + \partial_{x_2} u_2^\pm  = 0 \quad \text{ in  } T^\pm,  \text{ respectively. }
	\end{align*}
	\item \label{itm:bc}
	$\uu$ satisfies homogeneous Dirichlet boundary conditions on $\partial \Omega$.
	In combination with the fact that the gradients are piecewise continuous, we have
	\begin{align*}
		\nabla \uu^\pm (\xx) w^\pm = 0 \quad \text{ for } \xx \in e^\pm.
	\end{align*}
	\item \label{itm:cont}
	 $\uu$ is continuous along $e$, and with $\uu$ piecewise $C^1$ this implies that
in both components $i \in \{1,2\}$ the directional derivative in direction $e$
is continuous.
	 This means that
	\begin{align*}
		\nabla u_i^+(\xx)  \begin{pmatrix}  0 \\ 1
			\end{pmatrix} =	\nabla u_i^-(\xx)  \begin{pmatrix}  0 \\ 1
		\end{pmatrix} \qquad \text{ for } \xx \in e,\;i=1,2.
	\end{align*}
\end{enumerate}
	
	Since $\nabla \uu^\pm$ can be continuously extended to the boundary of $B^\pm$,
respectively, the identities hold in particular for $\xx = \bfz=(0,0)^t$.
	Writing the (unknown) vector of the partial derivatives
$\partial_{x_i} u_j^\pm(0)$ as a vector $D \in \mathbb{R}^8$, that is
	\begin{align*}
		D &= (	\partial_{x_1} u_1^+(0),	\partial_{x_2} u_1^+(0), 	\partial_{x_1} u_2^+(0), \partial_{x_2} u_2^+(0),...\\
		&\qquad 	\partial_{x_1} u_1^-(0),	\partial_{x_2} u_1^-(0), 	\partial_{x_1} u_2^-(0), \partial_{x_2} u_2^-(0) )^t
	\end{align*}
	we can formulate the above conditions as linear system of equations
	\begin{equation*}
	\underbrace{	\left(\begin{array}{cccc|cccc}
			1& 0 & 0  &1 & 0& 0&0&0\\
			1 & \alpha^+ & 0 & 0 &0&0 &0& 0 \\
			0 & 0 & 1 & \alpha^+ &0&0&0& 0\\
			0 & 1 & 0 & 0 & 0 & 1 & 0 & 0\\
			\hline
			0 & 0 & 0 & 1 & 0 & 0 & 0 & 1\\
			0& 0&0&0&	1& 0 & 0  &1  \\
			0 &0&0 &0 & -1 & \alpha^- & 0 & 0 \\
			 0&0&0& 0&0 & 0 & -1 & \alpha^-
		\end{array}\right)}_{\eqqcolon A} D = \begin{pmatrix}
		0 \\ 0 \\ 0 \\ 0 \\	0 \\ 0 \\ 0 \\ 0
	\end{pmatrix}.
	\end{equation*}
	Here the rows 1 and 6 contain Fact \ref{itm:div}, the rows 2,3 and 7,8 represent
Fact \ref{itm:bc}, and the rows 4 and 5 encode Fact \ref{itm:cont}.
	Aside from the entries in lines 4 and 5 this matrix has block structure. 
	Note that the homogeneous solution $D = (0,0,0,0,0,0,0,0)^t \in \mathbb{R}^8$ is a solution.
	To prove that it is the only solution, and hence we have that
$\nabla \uu(0) = 0$ at the boundary vertex $\bfz$, we investigate the matrix $A$ further.
	
	By symbolic calculation~\cite{matlabsymbolic} one can find that
	\begin{equation}\label{eqn:deteh}
		\det(A) = \alpha^- + \alpha^+.
	\end{equation}
Alternatively, starting from $A$ one can use the Laplace expansion of the determinant, applied twice, to rows 4 and 5 to compute the determinant of $A$.  
We can also derive \eqref{eqn:deteh} from the LU-decomposition of $A$
as $PA = LU$, where
$$ 
		P  =
					\left(\begin{array}{cccc|cccc}
	    1 &     0   &  0  &   0  &   0   &  0  &   0  &   0\\
		0  &   1  &   0   &  0  &   0  &   0  &   0  &   0\\
		0  &   0  &   1   &  0  &   0  &   0  &   0  &   0\\
		0  &   0  &   0   &  1  &   0  &   0  &   0  &   0\\
		0  &   0  &   0   &  0  &   0  &   1  &   0  &   0\\
		0  &   0  &   0   &  0  &   1  &   0  &   0  &   0\\
		0  &   0  &   0   &  0  &   0  &   0  &   0  &   1\\
		0  &   0  &   0   &  0  &   0  &   0  &   1  &   0
			\end{array}\right),  \qquad 
		 L =
			\left(\begin{array}{cccc|cccc}
		1&    0& 0&  0& 0&      0& 0& 0\\
		1&    1& 0&  0& 0&      0& 0& 0\\
		0&    0& 1&  0& 0&      0& 0& 0\\
		0& \frac{1}{\alpha^+}& 0&  1& 0&      0& 0& 0\\
		\hline
		0&    0& 0&  0& 1&      0& 0& 0\\
		0&    0& 0& \alpha^+& 0&      1& 0& 0\\
				0&    0& 0&  0& 0&      0& 1& 0 \\
		0&    0& 0&  0& -1& -\frac{\alpha^-}{\alpha^+} & 0& 1
	\end{array}\right),
$$
$$
\text{and} \;
		U = 	\left(\begin{array}{cccc|cccc}
		1&  0& 0&    1& 0&   0& 0&         0\\
		0& \alpha^+& 0&   -1& 0&   0& 0&         0\\
		0&  0& 1&   \alpha^+& 0&   0& 0&         0\\
		0&  0& 0& \frac{1}{\alpha^+}& 0&   1& 0&         0\\
		\hline
		0&  0& 0&    0& 1&   0& 0&         1\\
		0&  0& 0&    0& 0& - \alpha^+& 0&         1\\
		0&  0& 0&    0& 0&   0& -1&        \alpha^-\\
		0&  0& 0&    0& 0&   0& 0& \frac{\alpha^-}{\alpha^+} + 1
			\end{array}\right),
$$ 
provided that $\alpha^+ \neq 0$.
	Both $L$ and $U$ are triangular, so their determinants are the
product of their diagonal entries, and thus $\det(L)=1$ and
$\det(U) = \alpha^- + \alpha^+$.
	Since $P$ is a permutation matrix with two permutations,
$\det(P)=1$.
Thus we find
		$\det(A) = \det(U) = \alpha^- + \alpha^+$,
provided that $\alpha^+ \neq 0$.
This validates \eqref{eqn:deteh} on the dense set
$\{\alpha^+\neq 0\}$, and so it must hold everywhere.
Hence, the linear system of equations has the unique trivial solution if and only if
$w^+$ and $w^-$ are linearly independent.
This shows that, indeed, $\nabla \uu(\bfz) = \bfz$ is enforced,
if $w^\pm$ are linearly independent.

\bibliographystyle{plain}

\end{document}